\documentclass[letterpaper,12pt]{article} 

\usepackage{hyperref}
\usepackage[T1]{fontenc}
\usepackage[letterpaper,left=1.25in,right=1.25in,top=1in,bottom=1in]{geometry}
\usepackage{parskip}
\usepackage[english]{babel}
\usepackage{csquotes}
\usepackage[
backend=bibtex,
style=numeric,
sorting=nyt
]{biblatex}
\usepackage{microtype}
\usepackage{amsmath,amsthm,amssymb,amsfonts}
\usepackage{booktabs}
\usepackage{multirow}
\usepackage{xcolor}
\usepackage{graphicx}
\usepackage{subcaption}
\usepackage{tikz}
\usepackage{mathtools}
\usepackage{xparse}
\usepackage{placeins}
\usepackage[font=small,labelfont=bf]{caption}
\usetikzlibrary{positioning}
\usetikzlibrary{calc}

\DeclareMathOperator{\spec}{sp}

\usepackage{amssymb, amsthm, amsfonts, amsmath}
\usepackage{mathtools, mathrsfs}
\usepackage{xparse}
\usepackage{xcolor}
\makeatletter

\newcommand{\@bbset}[2]{\newcommand{#1}{\ensuremath{\mathbb{#2}}}}
\@bbset{\bbR}{R}
\@bbset{\bbQ}{Q}
\@bbset{\bbC}{C}
\@bbset{\bbZ}{Z}
\@bbset{\bbN}{N}
\@bbset{\bbF}{F}

\newcommand{\defeq}{\triangleq}

\DeclarePairedDelimiter\dparen{\lparen}{\rparen}

\DeclarePairedDelimiter\dbrack{\lbrack}{\rbrack}

\DeclarePairedDelimiter\norm{\lVert}{\rVert}

\DeclareMathOperator{\Exp}{E}

\makeatother

\title{Data-driven multiscale modeling for correcting dynamical systems}
\author{%
  Karl Otness, Laure Zanna \& Joan Bruna\\
  Courant Institute of Mathematical Sciences\\
  New York University\\
  \texttt{karl.otness@nyu.edu}
}
\date{}

\hypersetup{
  pdftitle={Data-driven multiscale modeling for correcting dynamical systems},
  pdfauthor={Karl Otness, Laure Zanna, and Joan Bruna},
}

\begin{document}

\maketitle{}

\begin{abstract}
  We propose a multiscale approach for predicting quantities in
  dynamical systems which is explicitly structured to extract
  information in both fine-to-coarse and coarse-to-fine directions. We
  envision this method being generally applicable to problems with
  significant self-similarity or in which the prediction task is
  challenging and where stability of a learned model's impact on the
  target dynamical system is important. We evaluate our approach on a
  climate subgrid parameterization task in which our multiscale
  networks correct chaotic underlying models to reflect the
  contributions of unresolved, fine-scale dynamics.
\end{abstract}

\section{Introduction}%
\label{sec:intro}

Typical deep learning methods across a wide range of application areas
make use of end-to-end learning. In such approaches, a neural network
is trained such that it receives feedback which matches the
requirements of the full task. However, in some applications this sort
of training is impractical or even impossible. Applications of
learning to scientific computing tasks---in particular to simulation
problems---frequently involve real-world dynamics which may not be
fully modeled or understood, or existing simulation software which is
difficult to integrate with learned models and which does not admit
backpropagation through simulation time steps. In these cases,
training uses strictly offline data (either pre-computed or derived
from real-world observations), while applying the network online after
training. As a result, the network may learn behaviors which are
successful on the offline training task, but which are unstable in
online evaluation~\cite{frezat22}.

In this work we examine an approach to decomposing prediction tasks
into separate prediction problems across scales, encouraging the
network to more completely exploit available multiscale information.
Networks making separate cross-scale predictions are trained
separately and \emph{not} end-to-end. We observe that this simple approach to
offline training improves model stability, and improves accuracy in
smaller architectures, allowing more efficient evaluation when
integrated into end applications. While we believe that our approach
may have wider applicability, we evaluate its performance on a set of
subgrid forcing prediction tasks which arise in real-world climate
modeling applications.

Climate models, which simulate the long-term evolution of the Earth's
atmosphere, oceans, and terrestrial weather, are critical tools for
projecting the impacts of climate change around the globe. Due to
limits on available computational resources, these models must be run
at a coarsened spatial resolution which cannot resolve all physical
processes relevant to the climate system~\cite{foxkemper14}. To
reflect the contribution of these subgrid-scale processes, closure
models are added to climate models to provide the needed subgrid-scale
forcing. These parameterizations model the contribution of these
fine-scale dynamics and are critical to high quality and accurate long
term predictions~\cite{ross23,foxkemper19}. A variety of approaches to
designing these parameterizations have been tested, ranging from
hand-designed formulations~\cite{smagorinsky63}, to modern machine
learning with genetic algorithms~\cite{ross23}, or neural networks
trained on collected
snapshots~\cite{zanna20,guillaumin21,maulik18,perezhogin23}, or in an
online fashion through the target simulation~\cite{frezat22}.

The problem of predicting these forcings is inherently multiscale; the
subgrid dynamics which must be restored represent the impact of the
subgrid and resolved scales on each other. Closure models for climate
are designed to be resolution-aware~\cite{jansen19}, but even so
existing deep learning subgrid models do not explicitly leverage the
interactions between scales, leaving it to the neural networks to
implicitly learn these relationships. Our approach makes these
interactions explicit.

\section{Approach}%
\label{sec:approach}

Recent methods for image generation have made use of diffusion
modeling where an image is sampled from a learned distribution over
several steps starting from noise, rather than training a network to
produce the result in one shot~\cite{yang22,song21}. One can view this
process as gradually filling in fine-scale details based on earlier
coarse-scale features. This approach has found a wide variety of
successful applications. Many simulation tasks, in which states must
be evolved in time, have dynamics which might benefit from this
approach---namely dynamics in which neighboring scales influence each
other and in which coarser-scale features may be easier to predict and
slower to evolve than finer-scale details.

While these approaches have been successful and provide a useful
inductive bias, this sampling method is generally quite expensive.
Many applications of machine learning to existing simulation tasks
insert a learned model as a \emph{component} of an existing simulation
and evaluate the learned model in each simulation step. In these
applications, further adding multiple diffusion steps may be
prohibitively costly. In this work, we try to keep many of the
benefits of this generative approach while reducing the cost of
evaluation by reducing the number of inference steps. More
specifically, we divide the state into two scale ranges: a ``high
resolution'' segment containing the fine scale details and a ``low
resolution'' segment containing only the coarse scales. Our prediction
process first produces a coarse, low resolution version of the target
field, then a second step uses this initial prediction to fill in the
fine scale details, yielding a high resolution output.

In particular, when predicting a field $S_{\mathsf{x}}$ we can try to
predict the quantity directly, depending on the current simulation
state $\mathsf{x}$. These quantities are often uncertain, but for our
purposes, we will use deterministic models to predict the expectation
of our target quantities. However, our approach could be extended to
stochastic models in several natural ways. A deterministic neural
network $f_{\theta}$ can be trained to perform this task directly:
\begin{equation}
  \label{eq:basepredict}
  f_{\theta}(\mathsf{x}) \approx \Exp\dbrack{S_{\mathsf{x}} | \mathsf{x}}
\end{equation}
In our multiscale approach we first predict a low resolution version
of the target field, $S_{\mathsf{x}\ \text{lr}}$ and condition on this
prediction while producing the full resolution output:
\begin{align}
  \label{eq:scalepredict}
  f_{\theta_1}^{\text{downscale}}(\mathsf{x}) &\approx \Exp\dbrack{S_{\mathsf{x}\ \text{lr}} | \mathsf{x}}\\
  \label{eq:scalepredict2}
  f_{\theta_2}^{\text{buildup}}(\mathsf{x}, S_{\mathsf{x}\ \text{lr}}) &\approx \Exp\dbrack{S_{\mathsf{x}} | \mathsf{x},\, S_{\mathsf{x}\ \text{lr}}}
\end{align}
We use the estimated expectation from
$f_{\theta_1}^{\text{downscale}}(\mathsf{x})$ in
Equation~\ref{eq:scalepredict} to provide a realization of
$S_{\mathsf{x}\ \text{lr}}$ in Equation~\ref{eq:scalepredict2}.
Further scale segments could be introduced if desirable for a
particular task. For our application, the low resolution field is
produced by low-pass filtering the full target, and resampling at a
lower resolution, yielding a field with smaller dimensions. That is,
$S_{\mathsf{x}\ \text{lr}} \defeq \mathcal{D} \circ
\mathcal{F}(S_{\mathsf{x}})$ for a low-pass spectral filter
$\mathcal{F}$ and a resampling operation $\mathcal{D}$.

To evaluate our multiscale approach, we consider the problem of
learning subgrid forcings for fluid models, a problem which arises in
climate modeling applications. In particular, we use two idealized
simulations: (1) a two layer quasi-geostrophic model \emph{QG}, and
(2) Kolmogorov flow \emph{KF}. These two models will be introduced in
further detail below and in Appendix~\ref{sec:modelextdef}. Each
target simulation autonomously evolves a set of state variables
through time and can be evaluated with a configurable grid resolution.
We refer to a general state variable $\mathsf{x}$ in the following
introduction. In each system states may be ported to lower resolutions
by coarse-graining and filtering.

For each system we generate ground truth data by running the model at
a very high (``true'') resolution. This produces trajectories
$\mathsf{x}_{\text{true}}(t)$ and time derivatives
$\partial{}\mathsf{x}_{\text{true}}(t)/\partial{}t$. Next we generate
training data at a high resolution by applying a system-dependent
coarsening and filtering operator $C$ giving variables
$\bar{\mathsf{x}} \defeq C(\mathsf{x})$. Given nonlinearities in the
target simulations, this coarsening does not commute with the dynamics
of the models. To correct for this we must apply a subgrid forcing
term $S_{\mathsf{x}}$ to the evolution of each state variable:
\begin{equation}
  \label{eq:forcingdef}
  S_{\mathsf{x}} \defeq\overline{\frac{\partial{}\mathsf{x}}{\partial{}t}} - \frac{\partial{}\bar{\mathsf{x}}}{\partial{}t}.
\end{equation}
Note that formally the forcing $S_{\mathsf{x}}$ is a function of the
state $\mathsf{x}_{\text{true}}$. In a climate modeling application we
do not have access to this variable and so we train a model
$f_{\theta}(\bar{\mathsf{x}}) \approx S_{\mathsf{x}}$ which may be
stochastic.

\begin{figure}
  \centering
  \begin{tikzpicture}
    \node[inner sep=0pt, draw=black, thick] (bigq1) at (0,0) {\includegraphics[width=2cm]{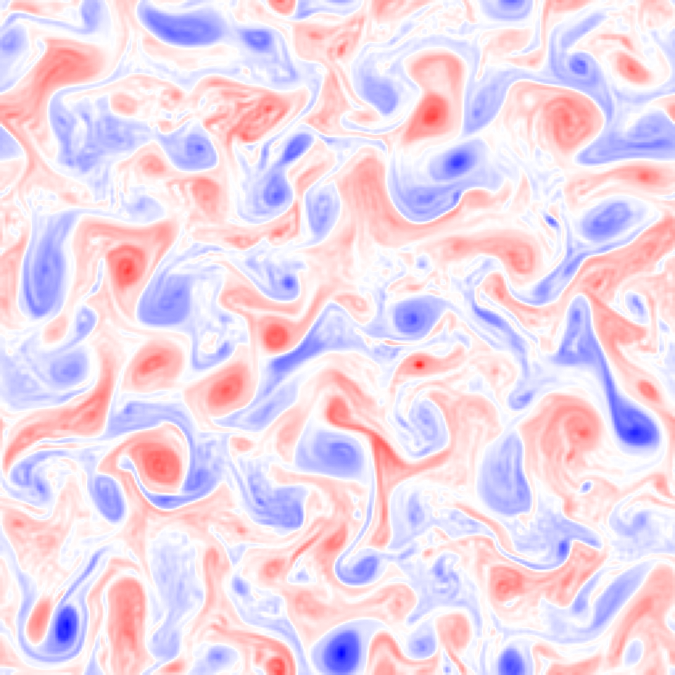}};
    \node[inner sep=0pt, draw=black, thick, right=2.5cm of bigq1] (bigq2) {\includegraphics[width=2cm]{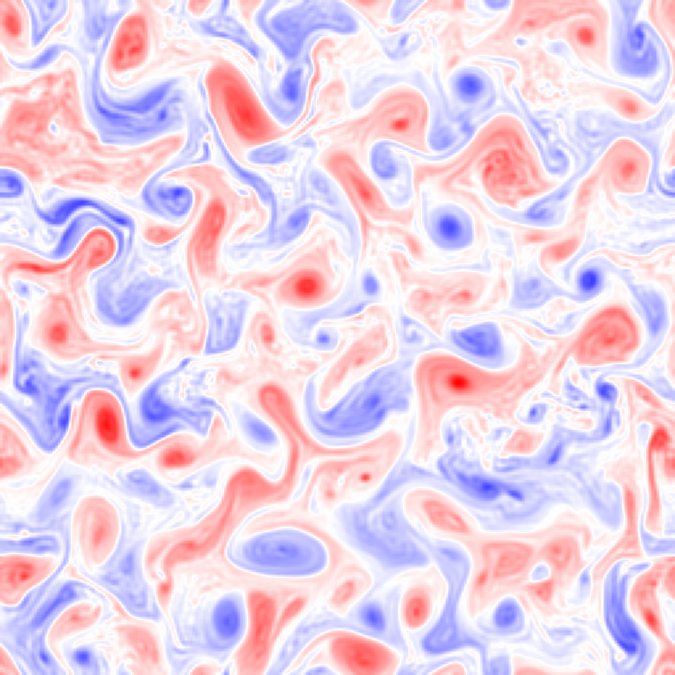}};
    \node[inner sep=0pt, draw=black, thick, right=2.5cm of bigq2] (bigq3) {\includegraphics[width=2cm]{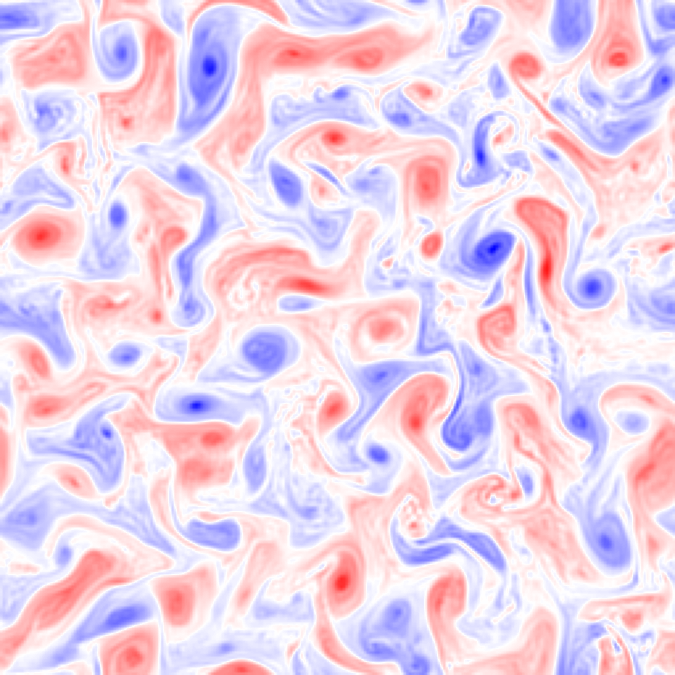}};

    \node[inner sep=0pt, draw=black, thick, below=of bigq1] (qhr1) {\includegraphics[width=1.65cm]{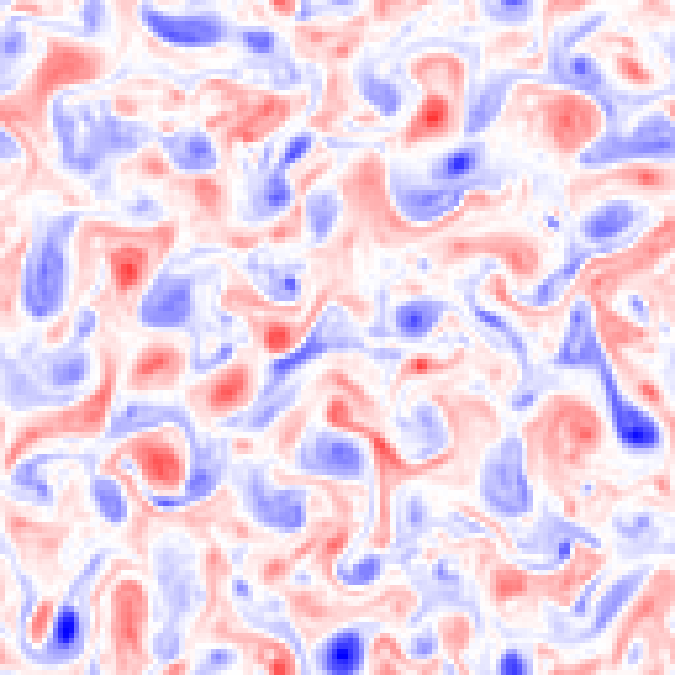}};
    \node[inner sep=0pt, draw=black, thick, right=0.15cm of qhr1] (shr1) {\includegraphics[width=1.65cm]{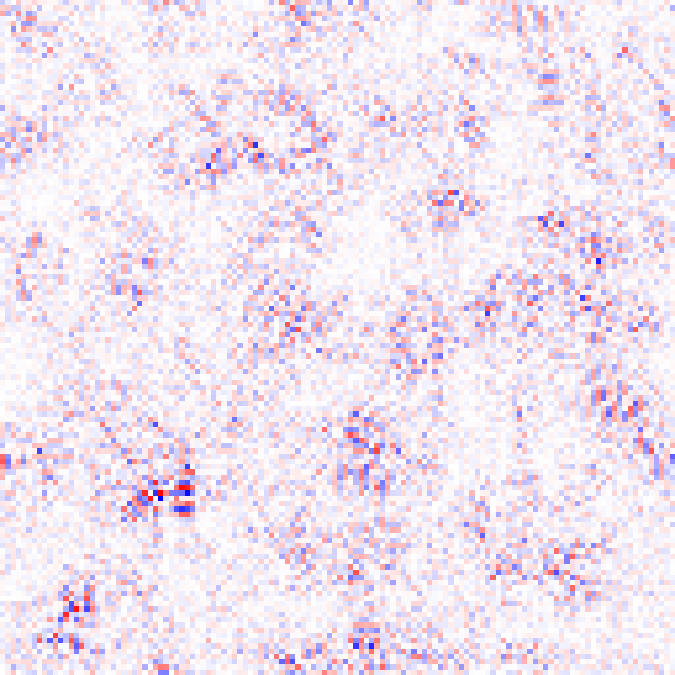}};
    \node[inner sep=0pt, draw=black, thick, below=of qhr1] (qlr1) {\includegraphics[width=1.25cm]{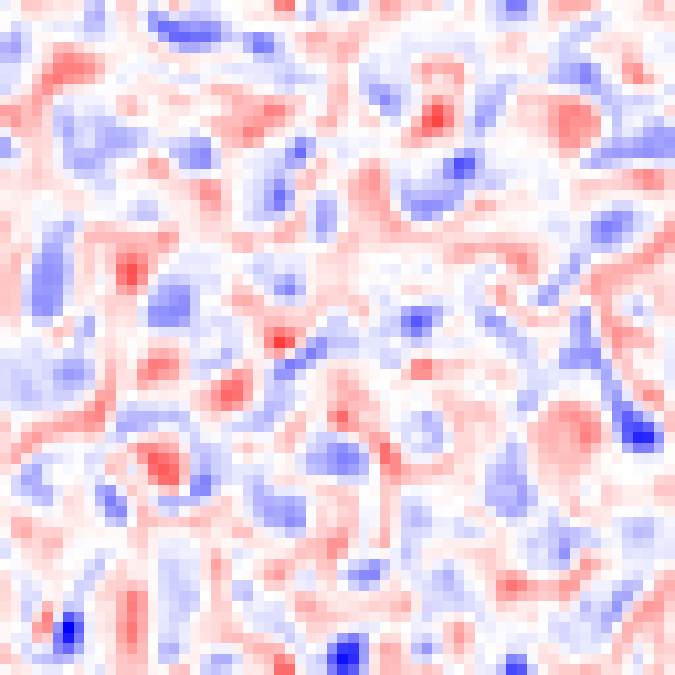}};
    \node[inner sep=0pt, draw=black, thick, below=of shr1] (slr1) {\includegraphics[width=1.25cm]{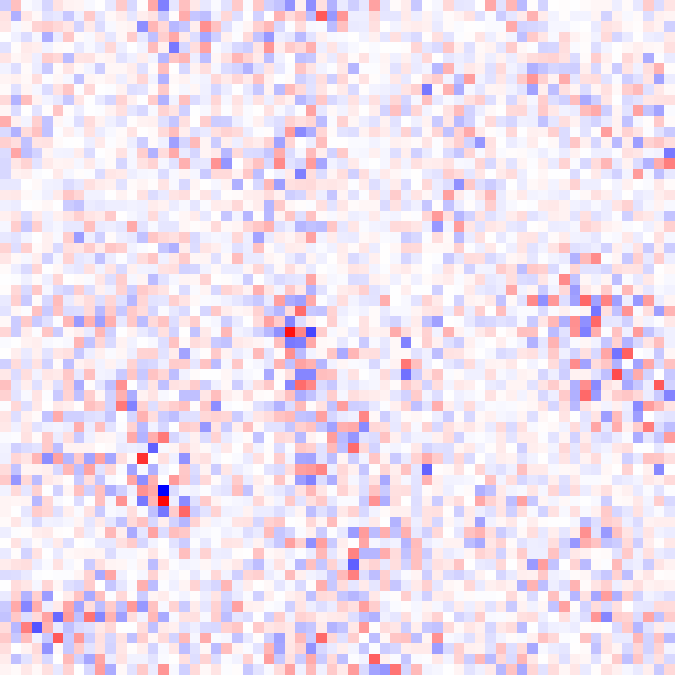}};

    \node[inner sep=0pt, draw=black, thick, below=of bigq2] (qhr2) {\includegraphics[width=1.65cm]{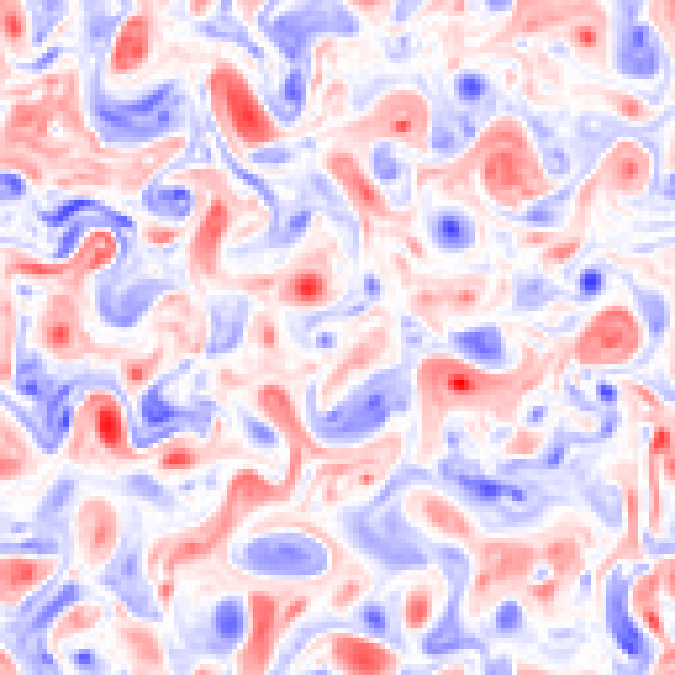}};
    \node[inner sep=0pt, draw=black, thick, right=0.15cm of qhr2] (shr2) {\includegraphics[width=1.65cm]{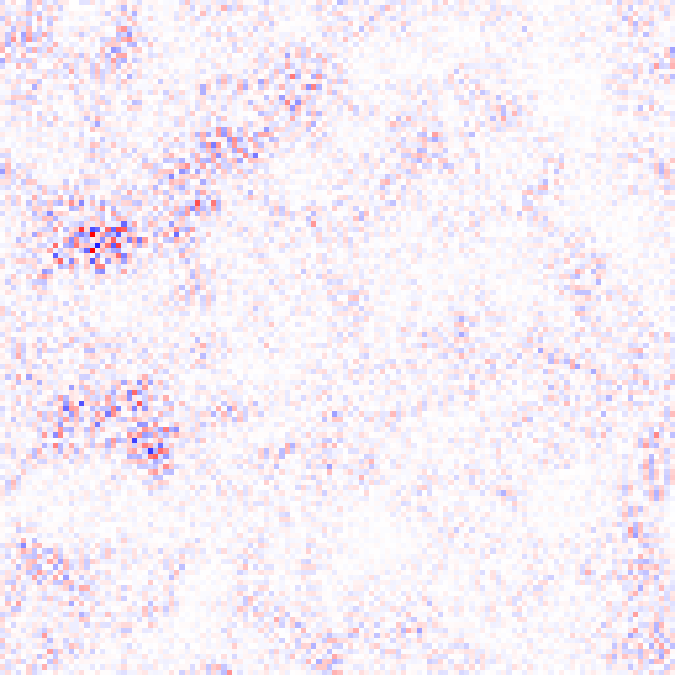}};
    \node[inner sep=0pt, draw=black, thick, below=of qhr2] (qlr2) {\includegraphics[width=1.25cm]{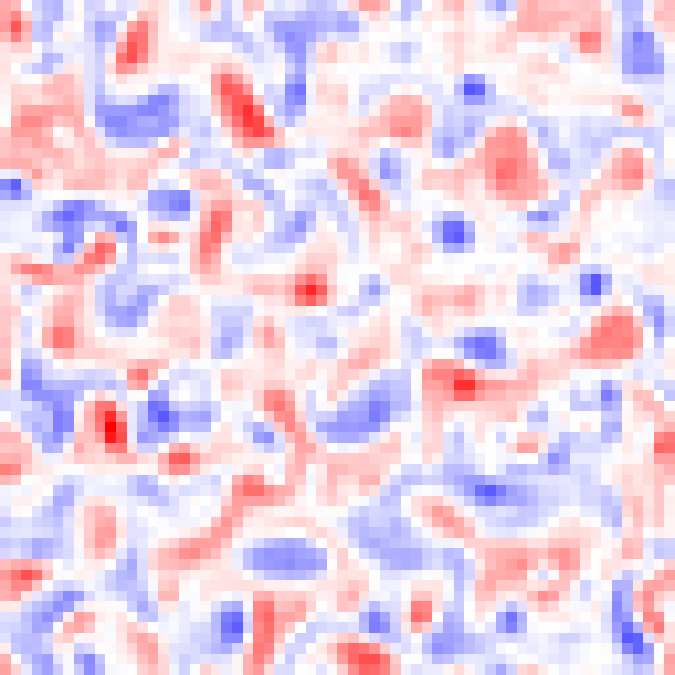}};
    \node[inner sep=0pt, draw=black, thick, below=of shr2] (slr2) {\includegraphics[width=1.25cm]{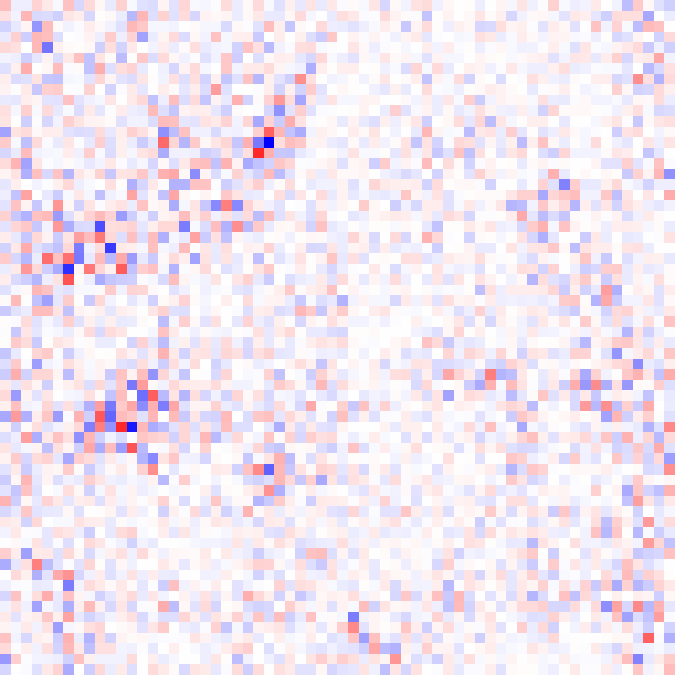}};

    \node[inner sep=0pt, draw=black, thick, below=of bigq3] (qhr3) {\includegraphics[width=1.65cm]{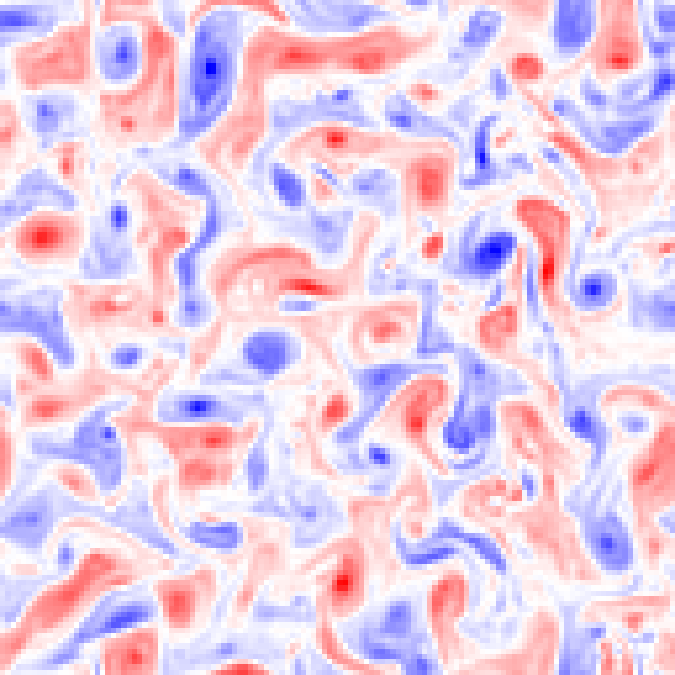}};
    \node[inner sep=0pt, draw=black, thick, right=0.15cm of qhr3] (shr3) {\includegraphics[width=1.65cm]{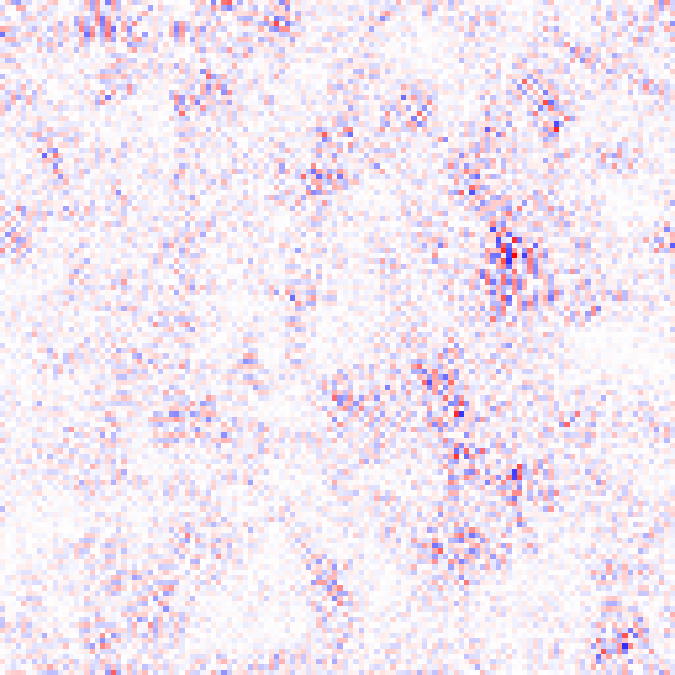}};
    \node[inner sep=0pt, draw=black, thick, below=of qhr3] (qlr3) {\includegraphics[width=1.25cm]{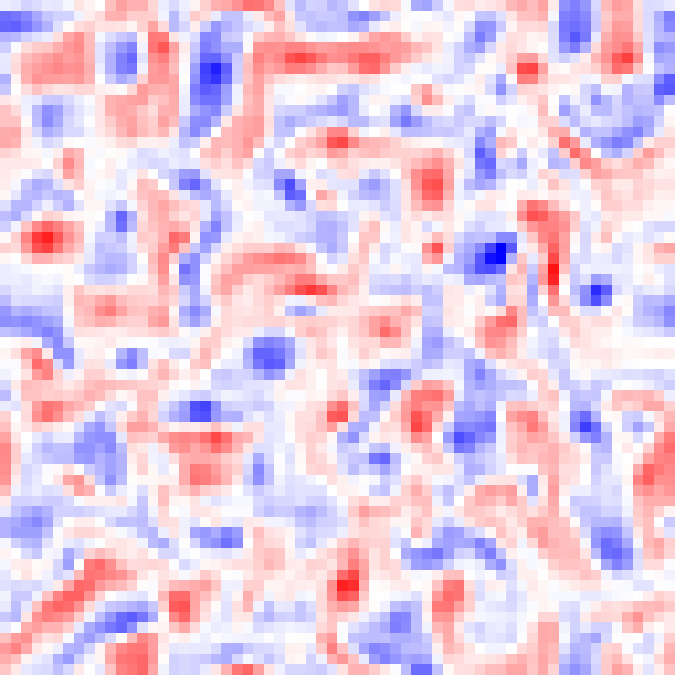}};
    \node[inner sep=0pt, draw=black, thick, below=of shr3] (slr3) {\includegraphics[width=1.25cm]{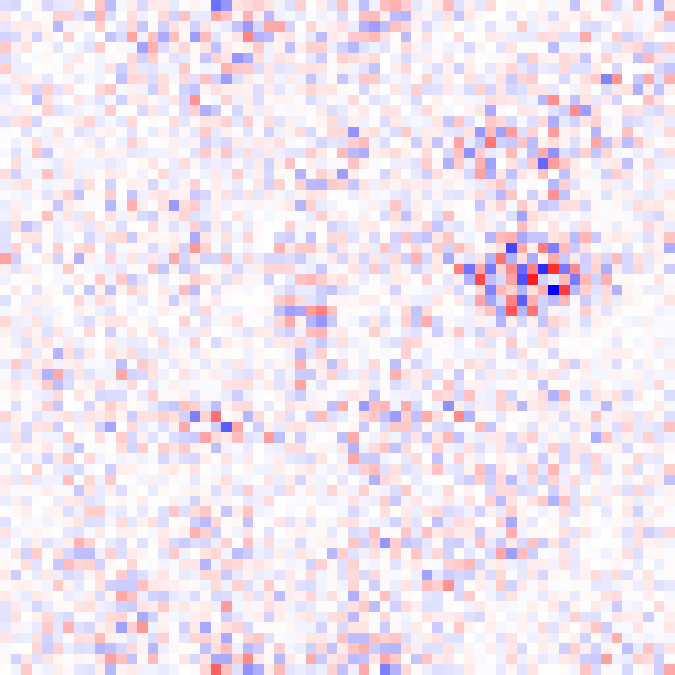}};

    \draw[-stealth] (bigq1) -- (bigq2) node[midway, above] {\includegraphics[width=0.6cm]{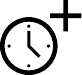}};
    \draw[-stealth] (bigq2) -- (bigq3) node[midway, above] {\includegraphics[width=0.6cm]{figures/ClockPlus.pdf}};

    \node (trueellip) at (shr3.north |- bigq3.east) {\large{}$\cdots$};

    \draw[-stealth, densely dashed] ([xshift=-0.2cm]qhr1.north east) to[out=50, in=130] node[midway, above] {\includegraphics[width=0.5cm]{figures/ClockPlus.pdf}} ([xshift=0.2cm]qhr2.north west);
    \draw[-stealth, densely dashed] ([xshift=-0.2cm]qhr2.north east) to[out=50, in=130] node[midway, above] {\includegraphics[width=0.5cm]{figures/ClockPlus.pdf}} ([xshift=0.2cm]qhr3.north west);

    \foreach \i in {1, ..., 3}
    {
      \node[rounded corners, inner sep=2pt, fill=white, fill opacity=0.7, text opacity=1, below right=2pt of bigq\i.north west] (lbigq\i) {$\mathsf{x}_{\text{true}}$};
      \node[rounded corners, inner sep=2pt, fill=white, fill opacity=0.7, text opacity=1, below right=2pt of qhr\i.north west] (lqhr\i) {$\mathsf{x}_{\text{hr}}$};
      \node[rounded corners, inner sep=2pt, fill=white, fill opacity=0.7, text opacity=1, below right=2pt of shr\i.north west] (lshr\i) {$S_{\mathsf{x}}$};
      \node[rounded corners, inner sep=2pt, fill=white, fill opacity=0.7, text opacity=1, below right=2pt of qlr\i.north west] (lqlr\i) {$\mathsf{x}_{\text{lr}}$};
      \node[rounded corners, inner sep=2pt, fill=white, fill opacity=0.7, text opacity=1, below right=2pt of slr\i.north west] (lslr\i) {$S_{\mathsf{x}\;\text{lr}}$};
    }
    \foreach \i in {1, ..., 3}
    {
      \draw[-stealth] (bigq\i) -- (qhr\i) node[midway, left] {$C$};
      \draw[-stealth] (qhr\i) -- (qlr\i) node[midway, left] {$D$};
      \draw[-stealth] (shr\i) -- (slr\i) node[midway, left] {$D$};
    }

    \node[left = 0.3cm of bigq1, align=center, font=\footnotesize] (sltrue) {True\\Resolution};
    \node[align=center, font=\footnotesize] (slhigh) at (sltrue.south |- qhr1.west) {High\\Resolution};
    \node[align=center, font=\footnotesize] (sllow) at (sltrue.south |- qlr1.west) {Low\\Resolution};
  \end{tikzpicture}
  \caption{The several scales involved in the experiments described below. The true resolution is used only while rolling out a ground truth reference trajectory and then discarded. Samples are resampled to lower resolutions using operators $C$ and $D$ described in Appendix~\ref{sec:coarsedef}. The values $S_{\mathsf{x}}$ are computed following Equation~\ref{eq:forcingdef} using both $\mathsf{x}_{\text{true}}$ and $\mathsf{x}_{\text{hr}}$. The high and low resolution samples are used for training. During evaluation the simulation is run online at the high resolution---following the dashed lines.}%
  \label{fig:scale-data-illust}
\end{figure}

We continue this process, introducing another downscaling%
\footnote{We use ``downscale'' and ``downscaling'' to refer to
  \emph{coarsening} a target variable, removing finer scales.}
operator $D$ and upscaling $D^{+}$. See Equation~\ref{eq:truncdef} for
the full definition. Taking
$\mathsf{x}_{\text{hr}} \defeq \bar{\mathsf{x}}$ as our high
resolution samples, we produce low resolution samples
$\mathsf{x}_{\text{lr}} \defeq D(\mathsf{x}_{\text{hr}})$ and
$S_{\mathsf{x}\;\text{lr}} \defeq D(S_{\mathsf{x}})$. This allows a
decomposition $\mathsf{x} = D^{+}D\mathsf{x} + \mathsf{x}'$ where
$\mathsf{x}'$ are the details removed by $D$. Our experiments thus
involve three resolutions, from fine to coarse: a ``true'' resolution;
a high resolution, $\text{hr}$; and a low resolution, $\text{lr}$. The
closures $S_{\mathsf{x}}$ try to update $\text{hr}$ to match the
``true'' resolution. The relationship between these scales and the
process of generating reference samples across them are illustrated in
Figure~\ref{fig:scale-data-illust}.

Just as predicting $S_{\mathsf{x}}$ from $\mathsf{x}_{\text{true}}$ is
fully deterministic, while predicting it from $\mathsf{x}_{\text{hr}}$
involves uncertainty, we anticipate a similar trend to hold for
$D(S_{\mathsf{x}})$. In other words, predicting $D(S_{\mathsf{x}})$
from $\mathsf{x}_{\text{hr}}$ should be easier than predicting
$D(S_{\mathsf{x}})$ directly from $\mathsf{x}_{\text{lr}}$. Then,
using this coarse-grained prediction $D(S_{\mathsf{x}})$ as a
foundation, we can learn to predict only the missing details and add
them. This process splits the problem of predicting $S_{\mathsf{x}}$
into two phases: (1) a ``downscale'' prediction to form
$D(S_{\mathsf{x}})$, and (2) a ``buildup'' prediction combining
$\mathsf{x}_{\text{hr}}$ and $D(S_{\mathsf{x}})$ to predict
$S_{\mathsf{x}}$, adding the missing details. This decomposition takes
advantage of self-similarity in the closure problem to pass
information between the coarse and fine scales and improve
predictions.

\section{Experiments}%
\label{sec:experiments}

To test this approach to predicting subgrid forcings we compare our
\emph{multiscale} approach against single-scale baselines. We select
the ``high'' resolution size in each target system (QG or KF) so that
the system requires closure (there are sufficient dynamics below the
grid-scale cutoff), but does not diverge~\cite{ross23}. An overview of
the two test systems is provided below with further details in
Appendix~\ref{sec:modelextdef}.

\subsection{Test Systems}%
\label{sec:test-systems}

We carry out our experiments on two systems: a two-layer
quasi-geostrophic system, and a single layer Kolmogorov Flow system
which is a configuration of Navier-Stokes.

\textbf{Quasi-Geostrophic System} This system is a two-layer
quasi-geostrophic model as implemented in PyQG and provides a
simplified approximation of fluid dynamics~\cite{pyqg}. For this work
we have produced a JAX port of this system\footnote{The ported QG model is available at \url{https://github.com/karlotness/pyqg-jax/}}~\cite{pyqgjax,jax}. This system
tracks the evolution of a potential vorticity $q$ divided into two
layers with periodic boundary conditions. Corrections $S_q$ are
applied to the time evolution of this field. Models receive the
current potential vorticity state $q$ as inputs and predict $S_q$
across both layers directly.

\textbf{Kolmogorov Flow System} We also test on a Kolmogorov Flow
system which is a single-layer incompressible Navier-Stokes flow with
periodic boundary conditions and a periodic forcing. We use an
implementation from JAX-CFD configured to have Reynolds number
7\,000~\cite{jaxcfd}. The state of this system is tracked by velocity
components $u$ and $v$ and all networks are also provided a computed
vorticity for each state $\omega$. Networks predict output quantities
$S_u$ and $S_v$, which are corrections applied to the velocity components.

Using these systems we carry out two sets of experiments: a small set
of ``separated'' experiments carried out on the QG model only (these
neural networks are trained and evaluated offline both with and
without access to the additional multiscale information); and
``combined'' experiments which run on both the QG and KF models which
provide an implementable closure model trained end to end. In both
cases we compare results against networks of equivalent architectures
without the additional multiscale structure.

For each experiment, a set of feedforward convolutional neural
networks is trained and evaluated separately. We train several
independently-initialized networks to capture the variance due to
initialization. These experiments also compare the performance of two
architectures, a ``small'' architecture and a ``large'' architecture
which are modifications of networks used in past
research~\cite{guillaumin21}. The small architecture for the combined
experiments was chosen as the result of an architecture search,
discussed below. Results are included in Section~\ref{sec:results},
and information on the network architectures and training procedure is
included in Appendix~\ref{sec:nndef}.

\subsection{Separated Experiments}%
\label{sec:sepexp}

We first carry out a set of preliminary tests on the quasi-geostrophic
system only. In these experiments, we examine the accuracy of the
learned forcings offline (that is, without rolling out trajectories)
while separating the two steps in our multiscale prediction process.
In particular, we examine the ability of networks to learn the
downscale and buildup prediction tasks, and check for the advantages
we intuitively expect from the additional multiscale information.

In these experiments, we train neural networks separately to predict
quantities between different scales. In particular we train
``downscale'' networks which predict only the low-resolution
components of the target forcing quantity while observing a high
resolution state, and ``buildup'' networks which work in the opposite
direction, predicting higher-resolution forcing details with access to
the low-resolution forcing. These illustrate some of the advantage
provided by the additional information and measure performance on
snapshots only (offline testing) as these separated networks do not
provide a fully-implementable closure model, since they require access
to an oracle to provide the additional input features.

Because these experiments target the QG system the target quantity is
the potential vorticity $q$. Each trained network receives a $q$ input
at the active (high resolution) simulation scale and predicts an $S_q$
output at the target scale.

\subsubsection{Downscale Prediction}%
\label{sec:downscale-prediction}

\begin{figure}
  \centering
  \begin{subfigure}[b]{0.45\linewidth}
    \centering
    \begin{tikzpicture}[node distance = 0.55cm and 1.2cm]
      \def\largesize{1.1cm}
      \def\smallsize{0.75cm}

      \node[inner sep=0pt, draw=black, thick] (q128) at (0, 0) {\includegraphics[width=\largesize]{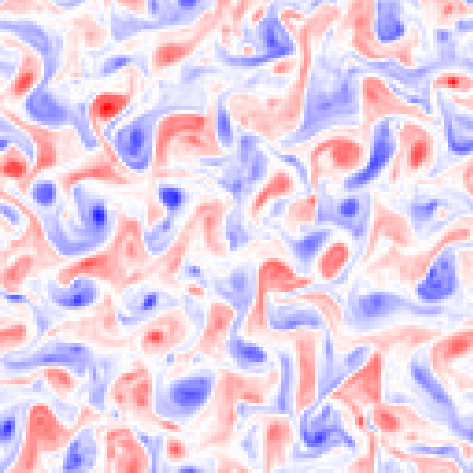}};
      \node[below left, color=darkgray] at (q128.north west) {$q_{\text{hr}}$};
      \node[inner sep=0pt, draw=black, thick, right=2cm of q128] (net-out) {\includegraphics[width=\largesize]{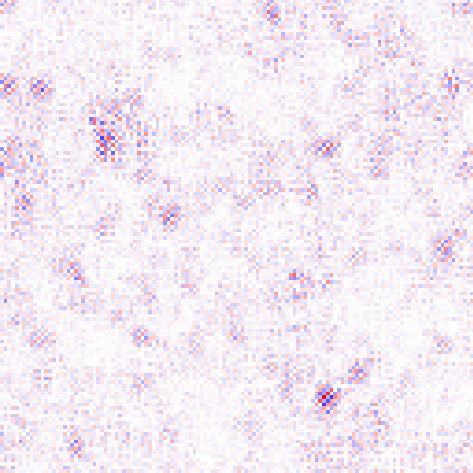}};
      \node[inner sep=0pt, draw=black, ultra thick, below=of net-out] (net-out-small) {\includegraphics[width=\smallsize]{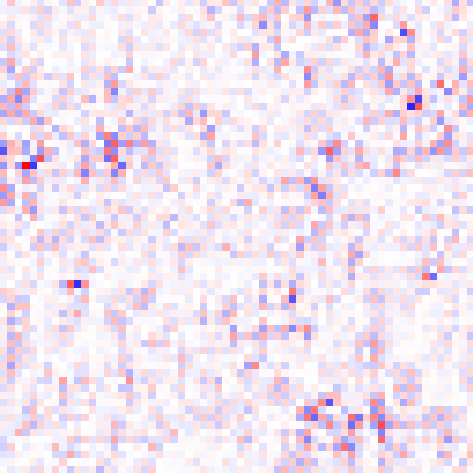}};
      \node[below right, color=darkgray] at (net-out-small.north east) {$\approx S_{\text{lr}}$};

      \draw[-stealth] (q128) -- (net-out) node[midway, above] {$f_{\theta}^{\text{downscale}}$};
      \draw[-stealth] (net-out) -- (net-out-small) node[midway, right] {$D$};
      \draw[-stealth, densely dashed] (q128) -- (net-out-small) node[midway, below left] {effective};
    \end{tikzpicture}
    \caption{Downscale prediction}%
    \label{fig:downscale}
  \end{subfigure}
  \begin{subfigure}[b]{0.45\linewidth}
    \centering
    \begin{tikzpicture}[node distance = 0.55cm and 1.2cm]
      \def\largesize{1.1cm}
      \def\smallsize{0.75cm}
      \node[inner sep=0pt, thick, draw=black] (q128) at (0, 0) {\includegraphics[width=\largesize]{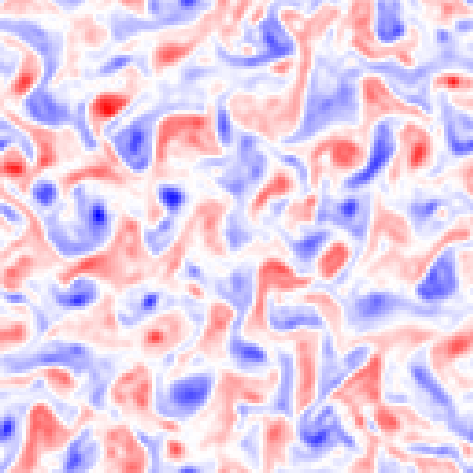}};
      \node[inner sep=0pt, thick, draw=black, below=of q128] (q64) {\includegraphics[width=\smallsize]{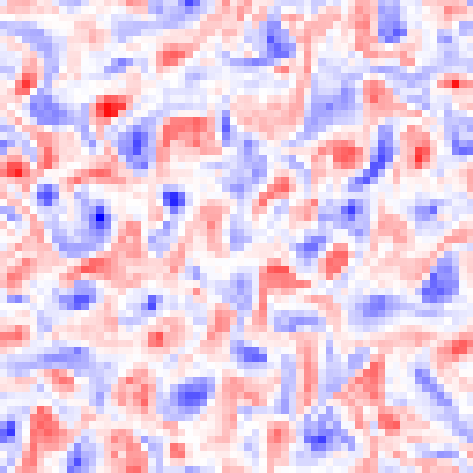}};
      \node[below left, color=darkgray] at (q64.north west) {$q_{\text{lr}}$};
      \node[inner sep=0pt, thick, draw=black, right=1.5cm of q128] (net-out) {\includegraphics[width=\largesize]{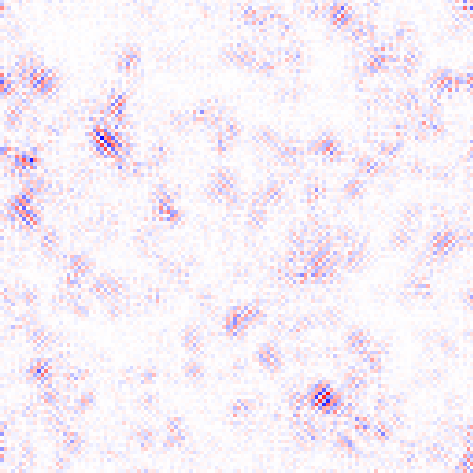}};
      \node[inner sep=0pt, ultra thick, draw=black, below=of net-out] (net-out-small) {\includegraphics[width=\smallsize]{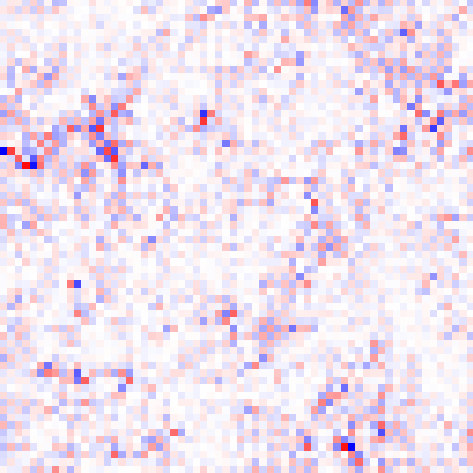}};
      \node[below right, color=darkgray] at (net-out-small.north east) {$\approx S_{\text{lr}}$};

      \draw[-stealth] (q64) -- (q128) node[midway, left] {$D^{+}$};
      \draw[-stealth] (q128) -- (net-out) node[midway, above] {$f_{\theta}^{\text{across}}$};
      \draw[-stealth] (net-out) -- (net-out-small) node[midway, right] {$D$};
      \draw[-stealth, densely dashed] (q64) -- (net-out-small) node[midway, above] {effective};
    \end{tikzpicture}
    \caption{Across prediction}%
    \label{fig:across}
  \end{subfigure}
  \caption{Downscale vs.\ across separated prediction tasks.
    The networks referenced in Equation~\ref{eq:datasks} are combinations of an inner network
    $f_{\theta}$ with the fixed rescaling operators $D$, $D^{+}$.
    The overall prediction is indicated with a dashed line.}%
  \label{fig:daillust}
\end{figure}

We compare the task of predicting $S_{\text{lr}} \defeq D(S_{\text{hr}})$ with
access to high resolution information $q_{\text{hr}}$ or restricted to
low resolution $q_{\text{lr}}$. This provides an estimate of the
advantage gained by predicting the target forcing with access to
details at a scale finer than that of the network's output. We train
two networks $f_{\theta}$ with the same architecture to perform one of two
prediction tasks:
\begin{equation}
  \label{eq:datasks}
  D \circ f_{\theta}^{\text{downscale}}(q_{\text{hr}}) \approx S_{\text{lr}} \qquad\text{and}\qquad D \circ f_{\theta}^{\text{across}} \circ D^{+}(q_{\text{lr}})\approx S_{\text{lr}}.
\end{equation}
To ensure that the convolution kernels process information at
the same spatial size, and differ only in the spectral scales
included, we first upsample all inputs to the same fixed size using a
spectral upscaling operator $D^{+}$ described in
Appendix~\ref{sec:coarsedef}.
The full prediction process including the re-sampling operators is
illustrated in Figure~\ref{fig:daillust}.

\subsubsection{Buildup Prediction}%
\label{sec:buildup-prediction}

We also test a prediction problem in the
opposite direction, predicting finer-scale details with access to
lower-resolution predictions, similar to a learned super-resolution
process used in recent generative modeling works~\cite{singer22,ho22}.
We train neural networks:
\begin{equation}
  \label{eq:bdtasks}
  f_{\theta}^{\text{buildup}}\dparen[\big]{q_{\text{hr}},\, D^{+}\dparen{S_{\text{lr}}}} \approx  S_{\text{hr}} - D^{+}(S_{\text{lr}}) \qquad\text{and}\qquad f_{\theta}^{\text{direct}}(q_{\text{hr}}) \approx S_{\text{hr}}\,,
\end{equation}
where $S_{\text{hr}} - D^{+}(S_{\text{lr}})$
are the details of $S_{\text{hr}}$ which are not reflected in
$S_{\text{lr}}$. The additional input
$S_{\text{lr}}$ is given by an oracle using ground truth data
in the training and evaluation sets.

\begin{figure}
  \centering
  \begin{subfigure}[b]{0.6\linewidth}
    \centering
    \begin{tikzpicture}[node distance = 0.55cm and 0.8cm]
      \def\largesize{1.1cm}
      \def\smallsize{0.75cm}

      \node[inner sep=0pt, draw=black, thick] (q128) at (0, 0) {\includegraphics[width=\largesize]{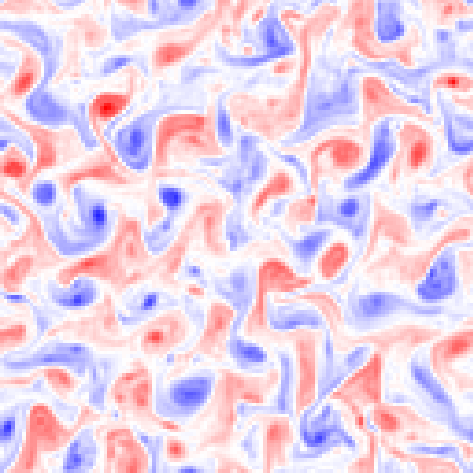}};
      \node[below left, color=darkgray] at (q128.north west) {$q_{\text{hr}}$};
      \node[inner sep=0pt, draw=black, below=0.25 of q128, thick] (s64-128) {\includegraphics[width=\largesize]{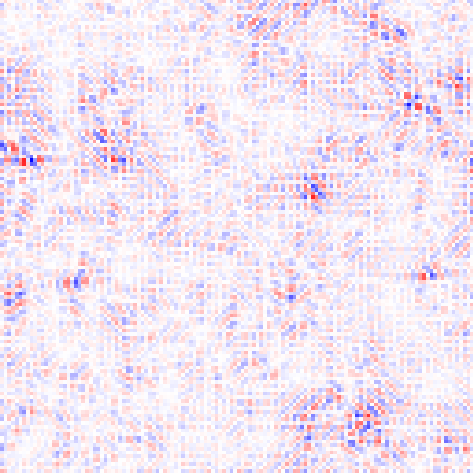}};
      \coordinate (midin) at ($(q128.east)!0.5!(s64-128.east)$);
      \node[inner sep=0pt, draw=black, right=1.65cm of midin, thick] (net-out) {\includegraphics[width=\largesize]{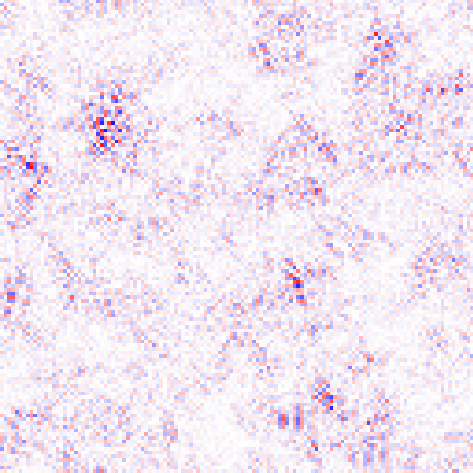}};
      \node[inner sep=0pt, draw=black, left=of s64-128, thick] (s64) {\includegraphics[width=\smallsize]{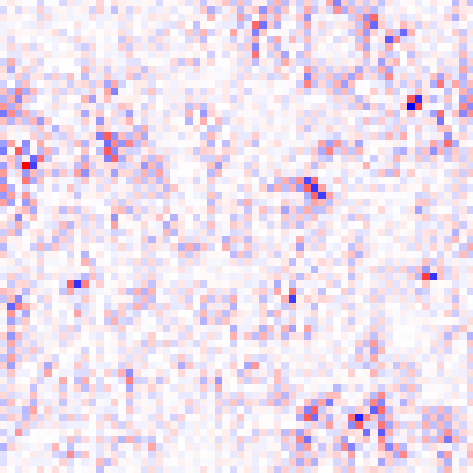}};
      \node[below left, color=darkgray] at (s64.north west) {$S_{\text{lr}}$};
      \coordinate (addy) at ($(s64-128.east)!0.5!(s64-128.south east)$);
      \coordinate (addx) at (net-out.south);
      \coordinate (addloc) at (addx |- addy);
      \node[draw, circle, inner sep=0pt] (add) at (addloc) {$+$};
      \node[inner sep=0pt, draw=black, right=3.25cm of s64-128, ultra thick] (net-out-added) {\includegraphics[width=\largesize]{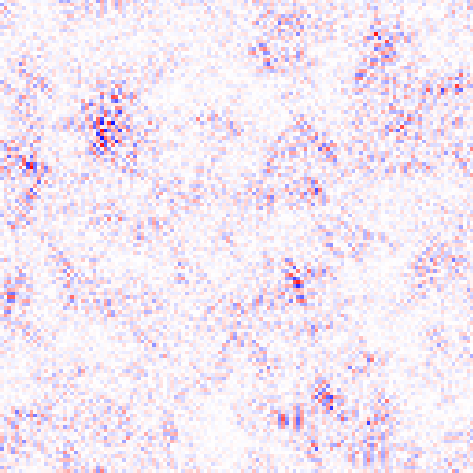}};

      \node[below right, color=darkgray] at (net-out-added.north east) {$\approx S_{\text{hr}}$};
      \coordinate (linejoin) at ($(midin)!0.3!(net-out.west)$);

      \draw[-stealth] (s64) -- (s64-128) node[midway, above] {$D^{+}$};
      \draw[-] (q128) -- (linejoin);
      \draw[-] (s64-128) -- (linejoin);
      \draw[-stealth] (linejoin) -- (net-out) node[near start, above, yshift=0.1cm] {$f_{\theta}^{\text{buildup}}$};
      \draw[-stealth] (net-out) -- (add);
      \draw[-stealth] (add) to[out=0, in=180] (net-out-added);
      \draw[-stealth] (s64-128) to[out=0, in=180] (add);

    \end{tikzpicture}
    \caption{Buildup prediction}%
    \label{fig:buildup}
  \end{subfigure}
  \begin{subfigure}[b]{0.35\linewidth}
    \centering
    \begin{tikzpicture}[node distance = 0.55cm and 0.8cm]
      \def\largesize{1.1cm}
      \def\smallsize{0.75cm}

      \node[inner sep=0pt, draw=black, thick] (q128) at (0, 0) {\includegraphics[width=\largesize]{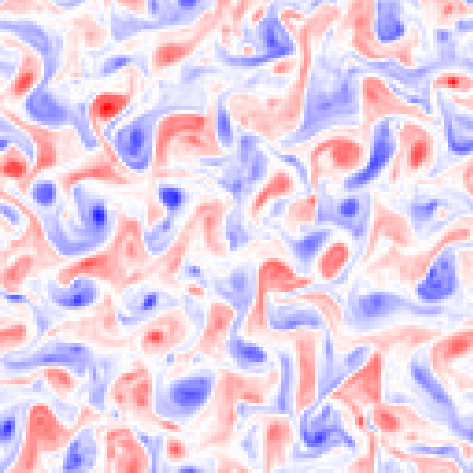}};
      \node[below left, color=darkgray] at (q128.north west) {$q_{\text{hr}}$};
      \node[inner sep=0pt, draw=black, right=1.5cm of q128, ultra thick] (net-out) {\includegraphics[width=\largesize]{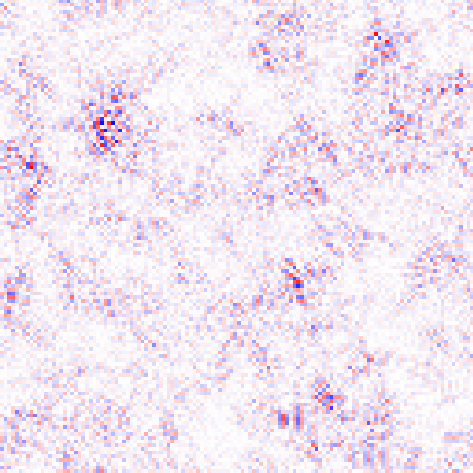}};
      \node[below right, color=darkgray] at (net-out.north east) {$\approx S_{\text{hr}}$};

      \draw[-stealth] (q128) -- (net-out) node[midway, above] {$f_{\theta}^{\text{direct}}$};
    \end{tikzpicture}
    \caption{Direct prediction}%
    \label{fig:direct}
  \end{subfigure}

  \caption{Buildup vs.\ direct separated prediction.
    The networks in Equation~\ref{eq:bdtasks} are
    combinations of the networks $f_{\theta}$ with the indicated fixed operations.
    In Figure~\ref{fig:buildup} $f_{\theta}$ predicts the details
    which are combined with $S_{\text{lr}}$
    from an oracle.}%
  \label{fig:bdillust}
\end{figure}

This experiment estimates the value in having a high-quality,
higher-confidence prediction $S_{\text{lr}}$, in addition to
$q_{\text{hr}}$, when predicting the details of $S_{\text{hr}}$. That
is, the experiment estimates the value in starting the prediction of
$S_{\text{hr}}$ by first locking in a coarse-grained version of the
target, and separately enhancing it with finer-scale features. The two
prediction tasks are illustrated in Figure~\ref{fig:bdillust}.

\subsection{Combined Experiments}%
\label{sec:combexp}

\begin{figure}
  \centering
  \begin{subfigure}{0.95\linewidth}
    \centering
    \begin{tikzpicture}[node distance = 0.55cm and 1.2cm]
      \def\largesize{1.1cm}
      \def\smallsize{0.75cm}

      \node[inner sep=0pt, draw=black, thick] (q128) at (0, 0) {\includegraphics[width=\largesize]{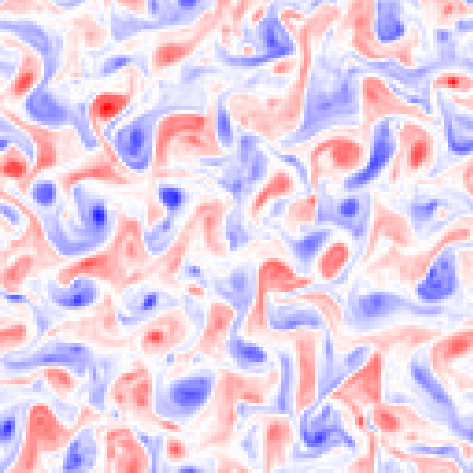}};
      \node[below left, color=darkgray] at (q128.north west) {$\mathsf{x}_{\text{hr}}$};
      \node[inner sep=0pt, draw=black, thick, right=2cm of q128] (net-out) {\includegraphics[width=\largesize]{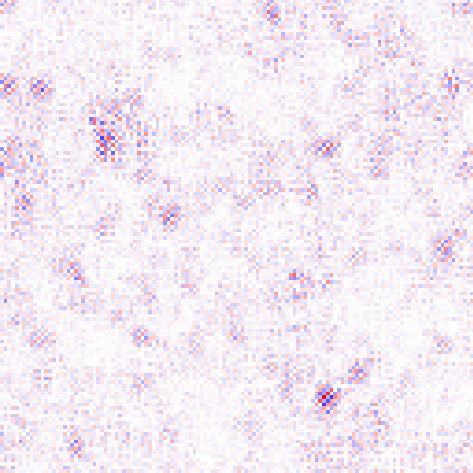}};
      \node[inner sep=0pt, draw=black, ultra thick, below=of net-out] (net-out-small) {\includegraphics[width=\smallsize]{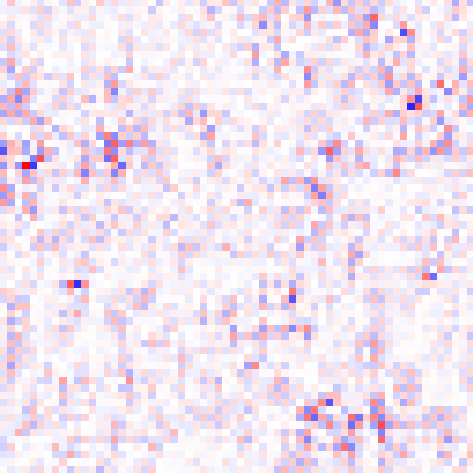}};
      \node[below left, color=darkgray] at (net-out-small.north west) {$S_{\mathsf{x}\ \text{lr}}\approx$};
      \draw[-stealth] (q128) -- (net-out) node[midway, below] {$f_{\theta_1}^{\text{downscale}}$};
      \draw[-stealth] (net-out) -- (net-out-small) node[midway, right] {$D$};

      \node[inner sep=0pt, draw=black, right=of net-out-small, thick] (s64-128) {\includegraphics[width=\largesize]{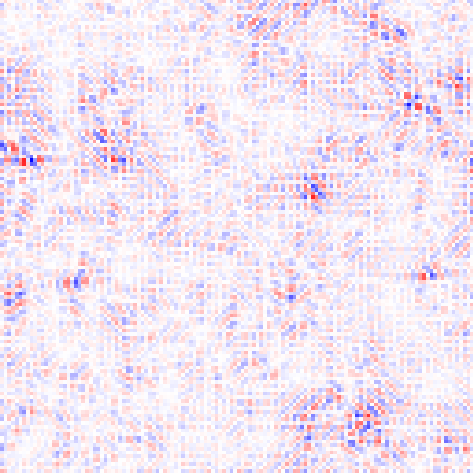}};
      \node[inner sep=0pt, draw=black, right=3.75cm of net-out, thick] (buildup-net-out) {\includegraphics[width=\largesize]{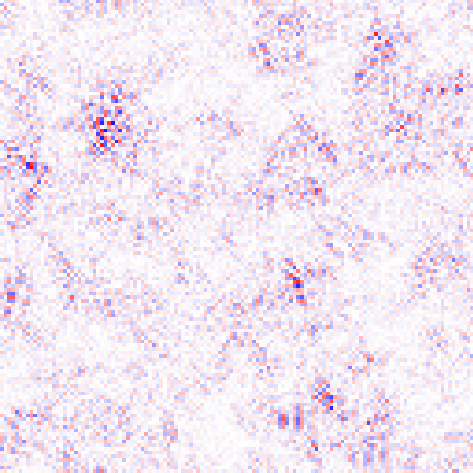}};
      \coordinate (addy) at (s64-128.east);
      \coordinate (addx) at (buildup-net-out.south);
      \coordinate (addloc) at (addx |- addy);
      \coordinate (midin) at ($(net-out.east)!0.75!(buildup-net-out.west)$);
      \node[draw, circle, inner sep=0pt] (add) at (addloc) {$+$};
      \node[inner sep=0pt, draw=black, right=of add, ultra thick] (net-out-added) {\includegraphics[width=\largesize]{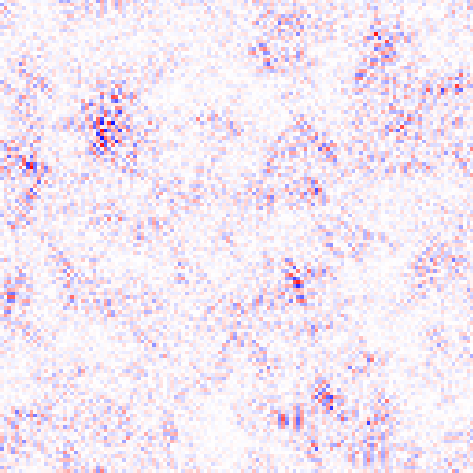}};

      \draw[-stealth] (net-out-small) -- (s64-128) node[midway, above] {$D^{+}$};
      \draw[-] (s64-128) to[out=35, in=205] (midin);
      \draw[-] (q128.north) to[out=15, in=155] (midin);
      \draw[-stealth] (midin) -- (buildup-net-out) node[near start, above, yshift=0.1cm] {$f_{\theta_2}^{\text{buildup}}$};
      \draw[-stealth] (buildup-net-out) -- (add);
      \draw[-stealth] (s64-128) -- (add);
      \draw[-stealth] (add) -- (net-out-added);
      \node[below right, color=darkgray] at (net-out-added.north east) {$\approx S_{\mathsf{x}\ \text{hr}}$};
    \end{tikzpicture}
    \caption{Full combined prediction flow}%
    \label{fig:comb-flow}
  \end{subfigure}

  \begin{subfigure}{0.95\linewidth}
    \centering
    \begin{tikzpicture}[node distance = 0.55cm and 1.2cm]
      \def\largesize{1.1cm}
      \def\smallsize{0.75cm}

      \node[inner sep=0pt, draw=black, thick] (q128) at (0, 0) {\includegraphics[width=\largesize]{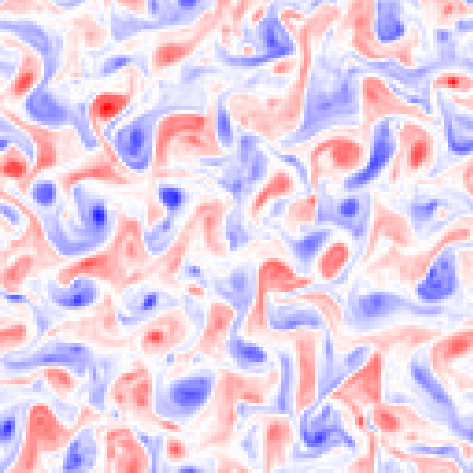}};
      \node[below left, color=darkgray] at (q128.north west) {$\mathsf{x}_{\text{hr}}$};
      \node[inner sep=0pt, draw=black, thick, right=2cm of q128] (net-out) {\includegraphics[width=\largesize]{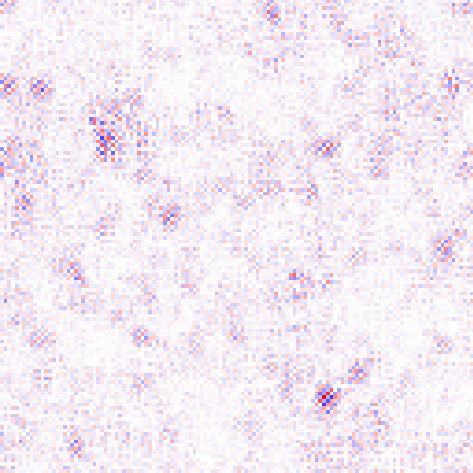}};
      \node[inner sep=0pt, draw=black, right=2cm of net-out, thick] (buildup-net-out) {\includegraphics[width=\largesize]{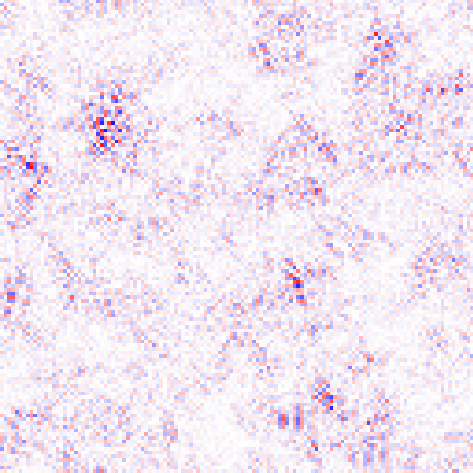}};
      \node[draw, circle, inner sep=0pt, below=of buildup-net-out] (add) {$+$};
      \node[inner sep=0pt, draw=black, right=of add, ultra thick] (net-out-added) {\includegraphics[width=\largesize]{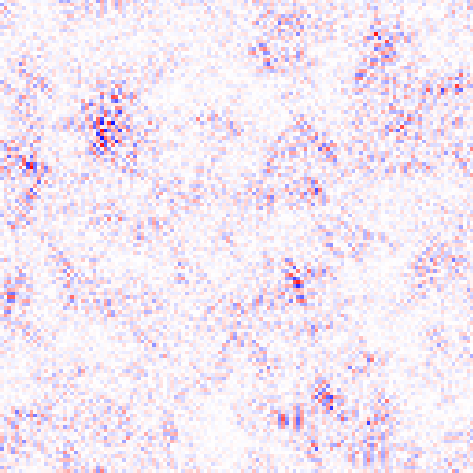}};

      \coordinate (midin) at ($(net-out.east)!0.75!(buildup-net-out.west)$);

      \draw[-stealth] (q128) -- (net-out) node[midway, below] {$f_{\theta_1}^{\text{step 1}}$};
      \draw[-] (q128.north) .. controls ++(18:3.5) and ++(180:0.75) ..  (midin);
      \draw[-stealth] (net-out) -- (buildup-net-out) node[midway, below] {$f_{\theta_2}^{\text{step 2}}$};
      \draw[-stealth] (buildup-net-out) -- (add);
      \draw[-stealth] (net-out) to[out=305, in=180] (add);
      \draw[-stealth] (add) -- (net-out-added);
      \node[below right, color=darkgray] at (net-out-added.north east) {$\approx S_{\mathsf{x}\ \text{hr}}$};
    \end{tikzpicture}
    \caption{Structure of the baseline for the combined prediction
      task}%
    \label{fig:comb-base}
  \end{subfigure}
  \caption{Flow for the combined prediction experiments. For the full
    combined flow each network $f_{\theta}$ is trained sequentially,
    and earlier weights are frozen and used to train networks later in
    the pipeline. The overall flow no longer requires an oracle for
    any additional inputs. Note also that the networks in the baseline
    flow in Figure~\ref{fig:comb-base} have the same residual
    prediction structure. The baseline network is trained end-to-end
    and differs only in that it misses the additional supervision from
    the multiscale training. The variable $\mathsf{x}$ represents the
    differing scalar fields for the QG and KF systems.}%
  \label{fig:combillust}
\end{figure}

In these experiments, we combine the networks trained in the
``downscale'' and ``buildup'' experiments, passing the downscale
prediction as an input to the buildup network. This removes the oracle
providing lower resolution predictions used to train the separate
networks. In each test, we choose two scale levels and first predict a
coarsened version of the subgrid forcing at the lowest resolution,
then gradually enhance it with missing scales using the buildup
process discussed above. These tests are carried out for both the
quasi-geostrophic and Kolmogorov flow systems. The overall flow of
this combined approach is illustrated in Figure~\ref{fig:combillust}
along with the associated baseline architecture.

Because this configuration yields an implementable closure model, we
perform our evaluations \emph{online}. That is, while the networks are
trained on snapshots, we evaluate the accuracy and stability of the
forcing by rolling out multiple trajectories using the trained
networks. As in the separated experiments we consider two network
architectures, a ``large'' architecture based on other works and a
``small'' architecture with fewer layers and smaller convolution
kernels. The small architecture was based on the results of an
architecture search discussed in Appendix~\ref{sec:nndef}. This allows
us to compare how architectural capacity may affect the behavior and
benefits of our multiscale approach.

For these experiments we retrain new neural networks building out the
training pipeline sequentially. We first train the first, downscale,
network, and then fix its weights and use its outputs to train the
subsequent buildup network. In this way, later networks see realistic
inputs during training rather than unrealistically clean data from a
training set oracle.

For a combined prediction across two scales $\text{lr}$ and
$\text{hr}$, we predict $S_{\mathsf{x}\ \text{hr}}$ from only $\mathsf{x}_{\text{hr}}$
following the procedure below:
\begin{align}
  \label{eq:combined-def}
  \begin{split}
    \tilde{S}_{\mathsf{x}\ \text{lr}} &\defeq D \circ f_{\theta_1}^{\text{downscale}}(\mathsf{x}_{\text{hr}}) \approx S_{\mathsf{x}\ \text{lr}}\\
    S_{\mathsf{x}\ \text{hr}} &\approx f_{\theta_2}^{\text{buildup}}\dparen[\big]{\mathsf{x}_{\text{hr}},\, D^{+}(\tilde{S}_{\mathsf{x}\ \text{lr}})} + D^{+}(\tilde{S}_{\mathsf{x}\ \text{lr}}).
  \end{split}
\end{align}
The quantity $\tilde{S}_{\mathsf{x}\ \text{lr}}$ is an approximate
neural network output used in subsequent predictions.
Equation~\ref{eq:combined-def} composes the prediction tasks described
in Equation~\ref{eq:datasks} and Equation~\ref{eq:bdtasks}. See
Figure~\ref{fig:comb-flow} for an illustration of the above flow.
Each network $f_{\theta}$ is trained separately
against either $S_{\mathsf{x}\ \text{lr}}$ or
$S_{\mathsf{x}\ \text{hr}}$ as appropriate.

The networks are evaluated on their ability to produce a stable
trajectory when rolled out, and their ability to correct the energy
spectrum---adding and removing energy across spectral scales as needed
to correct issues from coarse-graining. For these experiments, we examine the
trends in the system's total kinetic energy across time and
distributions in the spectral error of the system, computed by adding
errors in the average kinetic energy spectrum of a trajectory. These
measurements are distinct from those used to compare the quality of
snapshots used for the offline tests conducted as part of the
separated experiments.

\section{Results}%
\label{sec:results}

In this section we describe the results of our experiments and
measurements. Details of the approach taken in each experiment are
provided in Section~\ref{sec:experiments}.

\subsection{Separated Experiments}%
\label{sec:results-sepexp}

For both the downscale and buildup prediction tasks, we train three
neural networks. Once trained, we evaluate their performance on a
held out evaluation set measuring performance with three metrics: a
mean squared error (MSE), a relative $\ell_2$ loss, and a relative
$\ell_2$ of the spectra of the predictions.

The MSE is a standard mean squared error evaluated over each sample
and averaged. The other two metrics are derived from previous work
evaluating neural network parameterizations~\cite{perezhogin23} (where they
were called $\mathcal{L}_{\text{rmse}}$ and $\mathcal{L}_{\text{S}}$).
The metrics in this previous work were originally designed to measure performance for stochastic
subgrid forcings. Here we use the two metrics from that work which do
not collapse to trivial results for deterministic models.
These are defined as:
\begin{equation}
  \label{eq:metricdef}
  \text{Rel $\ell_2$} \defeq \frac{\norm{S - \tilde{S}}_2}{\norm{S}} \qquad\text{and}\qquad \text{Rel Spec $\ell_2$} \defeq \frac{\norm{\spec(S) - \spec(\tilde{S})}_2}{\norm{\spec(S)}_2}
\end{equation}
where $S$ is the true target forcing, $\tilde{S}$ is a neural network
prediction being evaluated, and $\spec$ is the isotropic power
spectrum. See \texttt{calc\_ispec} in PyQG for calculation details~\cite{pyqg}.
Each of these three metrics is averaged across the same
batch of 1024 samples selected at random from the set of held out
trajectories in the evaluation set.

Table~\ref{tab:higherres} shows the results for the downscale
experiments, comparing against ``across'' prediction which accesses
only coarse-scale information. In these results we observe an
advantage to performing the predictions with access to
higher-resolution data (the ``downscale'' columns), suggesting
potential advantages and a decrease in uncertainty in such predictions.

\begin{table}
  \centering
  {
    \small
    \begin{tabular}{llrrrrrr}
      \toprule
      \multirow{2}{*}{NN Size}                & \multirow{2}{*}{Metric}            & \multicolumn{2}{c}{$128\to96$} & \multicolumn{2}{c}{$128\to64$} & \multicolumn{2}{c}{$96\to64$}\\
      \cmidrule(r){3-4} \cmidrule(lr){5-6} \cmidrule(l){7-8}
                                              & & Downscale & Across & Downscale & Across & Downscale & Across \\
      \midrule
      \multirow{3}{*}{Small}  & MSE & 0.054&0.072&0.002&0.006&0.032&0.058\\
                                              & Rel $\ell_2$ & 0.317&0.364&0.394&0.629&0.348&0.469\\
                                              & Rel Spec $\ell_2$ & 0.133&0.145&0.154&0.471&0.154&0.254\\
      \addlinespace
      \multirow{3}{*}{Large}  & MSE & 0.038&0.057&0.002&0.006&0.024&0.052\\
                                              & Rel $\ell_2$ & 0.259&0.316&0.335&0.595&0.297&0.436\\
                                              & Rel Spec $\ell_2$ & 0.092&0.129&0.125&0.443&0.103&0.212\\
      \bottomrule
    \end{tabular}
  }
  \caption{Evaluation results for downscale vs.\ across generation. In
    all metrics, lower is better. The numbers in the first row of the
    table heading show the different scales involved in both
    prediction tasks. The results contributing to the MSE averages in
    this table are illustrated in Figure~\ref{fig:downscatter}.}%
  \label{tab:higherres}
\end{table}

Results for experiments examining prediction in the opposite
direction---predicting a high-resolution forcing with access to a
low-resolution copy of the target from an oracle---are included in
Table~\ref{tab:buildup}. We also observe an advantage in this task
from having access to the additional information. The low resolution
input in the buildup experiments yields lower errors on average at
evaluation. This advantage is greater when the additional input is
closer in scale to the target output. The predictions building up from
$96\times96$ to $128\times128$ have lower errors than those which
access an additional $64\times64$ input. This is not unexpected given
that the input with nearer resolution resolves more of the target
value, leaving fewer details which need to be predicted by the
network.

\begin{table}
  \centering
  {
    \small
    \begin{tabular}{llrrrrr}
      \toprule
      \multirow{2}{*}{NN Size} & \multirow{2}{*}{Metric} & Buildup & Buildup & Direct & Buildup & Direct\\
                               & & $64 \to 128$ & $96 \to 128$ & 128 & $64 \to 96$ & 96\\
      \midrule
      \multirow{3}{*}{Small}  & MSE & 0.094&0.033&0.097&0.060&0.108\\
                               & Rel $\ell_2$ & 0.314&0.187&0.319&0.251&0.333\\
                               & Rel Spec $\ell_2$ & 0.138&0.054&0.139&0.095&0.162\\
      \addlinespace
      \multirow{3}{*}{Large}  & MSE & 0.057&0.019&0.062&0.037&0.071\\
                               & Rel $\ell_2$ & 0.242&0.141&0.251&0.195&0.268\\
                               & Rel Spec $\ell_2$ & 0.074&0.029&0.084&0.041&0.091\\
      \bottomrule
    \end{tabular}
  }
  \caption{Evaluation results from buildup vs.\ direct experiments. In
    all metrics, lower is better. The numbers in the second row of the
    table heading show the different scales involved in both
    prediction tasks. The results contributing to the MSE averages in
    this table are illustrated in Figure~\ref{fig:upscatter}.}%
  \label{tab:buildup}
\end{table}

\begin{figure}
  \centering
  \begin{subfigure}[t]{0.49\linewidth}
    \centering
    \includegraphics[width=\linewidth]{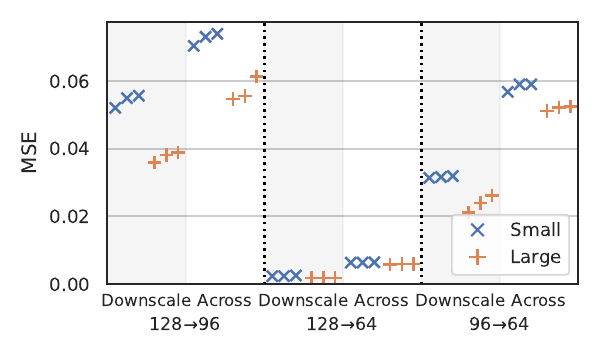}
    \caption{Downscale vs.\ across MSE}%
    \label{fig:downscatter}
  \end{subfigure}
  \begin{subfigure}[t]{0.49\linewidth}
    \centering
    \includegraphics[width=\linewidth]{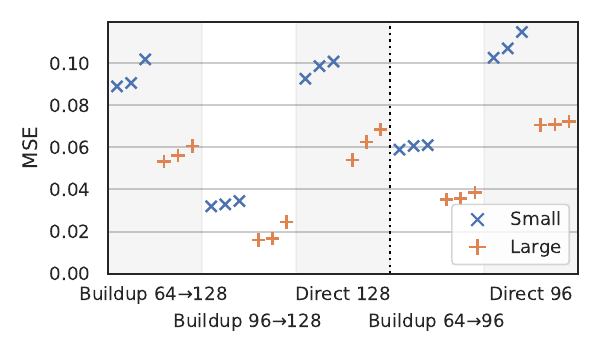}
    \caption{Buildup vs.\ direct MSE}%
    \label{fig:upscatter}
  \end{subfigure}
  \caption{Evaluation results from both of the separated experiments
    for the MSE metric. These are the same numbers which are reported
    as averages in Table~\ref{tab:higherres} and
    Table~\ref{tab:buildup}. The plot here shows the three
    samples---one from each trained network---used to compute the
    means.}%
  \label{fig:evalscatter}
\end{figure}

The results for both separated experiments (those reported in
Table~\ref{tab:higherres} and Table~\ref{tab:buildup}) for the MSE
metric are illustrated in Figure~\ref{fig:evalscatter}.

\subsection{Combined Experiments}%
\label{sec:results-combexp}

We also carry out tests comparing the performance of networks using
our multiscale prediction approach to a network predicting only at one
scale throughout. Figure~\ref{fig:combillust} illustrates the overall
flows of these networks. In the combined task, each independent
cross-scale network is trained separately in phases. Earlier networks
in the pipeline are trained and have their weights frozen and these
are used to produce inputs during the training of networks later in
the pipeline. The baseline for this configuration is trained
end-to-end and has the same residual structure and number of weights
as the multiscale network, but without any enforced predictions across
scales.

This setting produces a trained parameterization network which can be
applied to a real simulation and tested online. As a result, we
evaluate these networks by rolling out trajectories from a held out
test set. These are compared against a reference trajectory which was
originally produced at a ``true'' resolution. Because these systems
are chaotic we do not expect to reproduce exact states in these
trajectories and we do not examine snapshot errors over long time
horizons. Instead we compare statistical properties of these
trajectories including their total kinetic energy (which gives a sense
of stability), errors in the trajectory's energy spectrum (providing a
sense of the quality of the parameterization), and for the KF system,
the vorticity decorrelation time. Overall, we find that the multiscale
approach improves the stability of the learned parameterizations while
also permitting smaller neural network architectures to be used for
these tasks. Results of our experiments for the QG and KF systems are
described below.

\subsubsection{Quasi-Geostrophic Results}%
\label{sec:comb-qgres}

For the QG system we simulate the evolution of a held out trajectory
and conduct tests over four independently trained neural networks and
16 held out trajectories for each network. For these tests we are
concerned with two aspects: the stability of the system running with
the neural network closure, and the quality of the resulting
parameterization---in particular the extent to which it improves the
energy spectrum of the trajectory.

\begin{figure}
  \centering
  \begin{subfigure}[t]{0.9\linewidth}
    \centering
    \includegraphics[width=\linewidth]{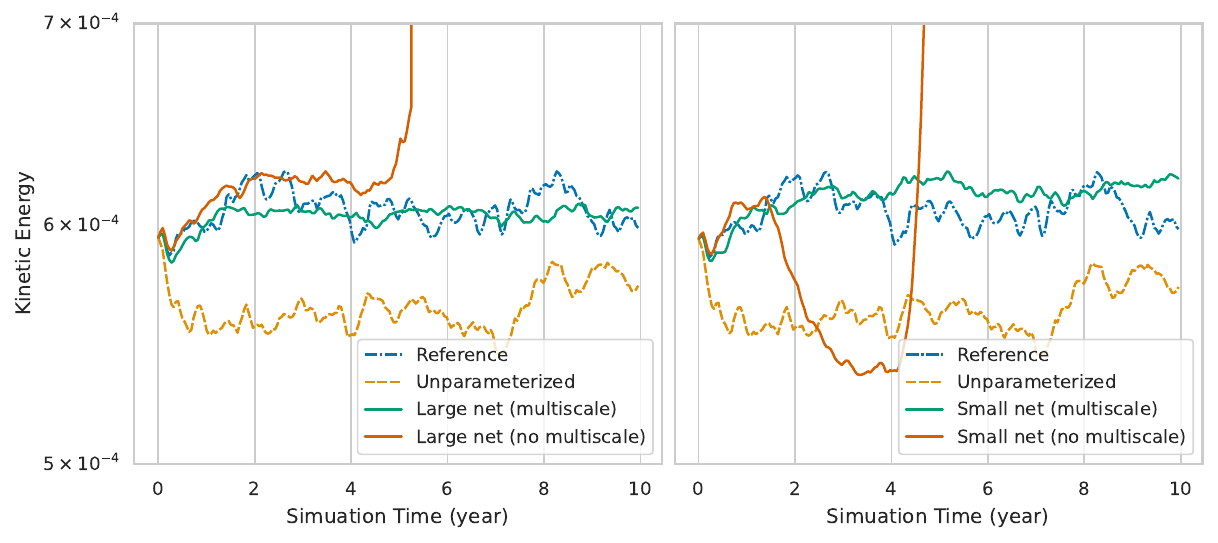}
    \caption{Scale 96---multiscale networks predict between $96\leftarrow64$}%
    \label{fig:qg-ke-time-96}
  \end{subfigure}
  \begin{subfigure}[t]{0.9\linewidth}
    \centering
    \includegraphics[width=\linewidth]{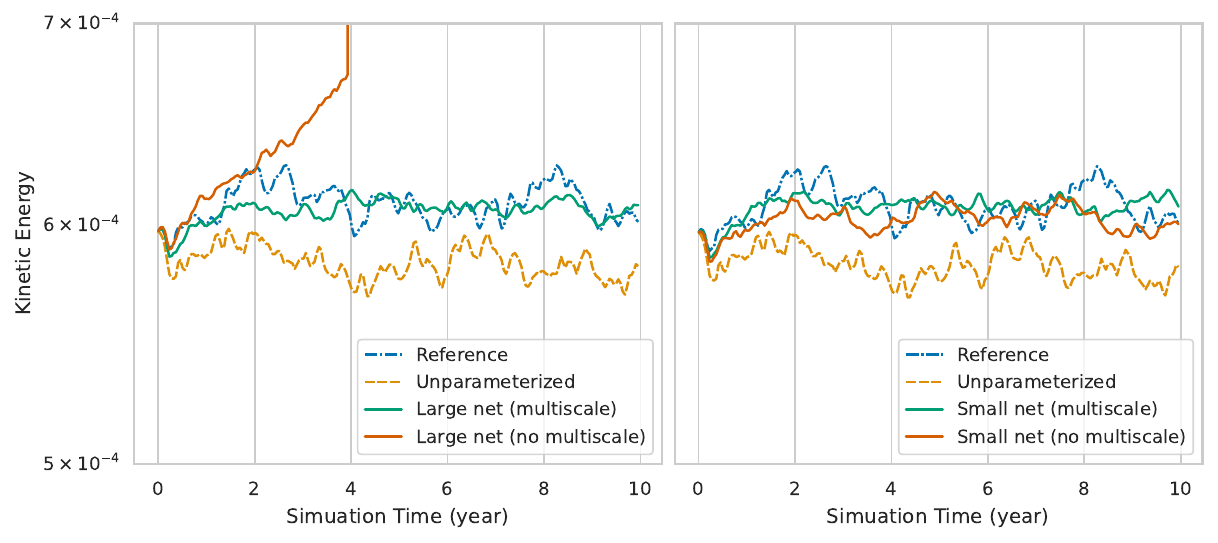}
    \caption{Scale 128---multiscale networks predict between $128\leftarrow96$}%
    \label{fig:qg-ke-time-128}
  \end{subfigure}
  \caption{Time evolution of mean kinetic energy for the QG system
    simulated at different scales. Trajectories are simulated for 16
    held out trajectories using 4 independently-trained networks.}%
  \label{fig:qg-ke-time}
\end{figure}

Figure~\ref{fig:qg-ke-time} shows variation in the mean kinetic energy
over time. We intend this as a rough measure of the stability of the
parameterized system providing a rough idea of whether the system is
under- or over-energized which can lead to instability and eventual
collapse in the system. The QG system is simulated at a grid size of
$96\times96$ and $128\times128$ with separate sets of neural networks
trained for each case. The simulations at a size of 96 have a greater
range of unrealized dynamics leading to a more challenging closure
problem. We note in particular that the trajectories which display
instability are those using networks without the multiscale component.
The smaller networks for the system at size 128 all display general
kinetic energy stability; however, with only kinetic energy statistics
this is difficult to distinguish from a network which applies no
parameterization at all. The QG system is stable without a
parameterization due to a fixed spectral filter which attenuates high
frequencies, and the target parameterization values have zero mean
which can in some cases lead a network to learn to apply no correction
which still yields a stable trajectory.

\begin{figure}
  \centering
  \begin{subfigure}[t]{0.9\linewidth}
    \centering
    \includegraphics[width=\linewidth]{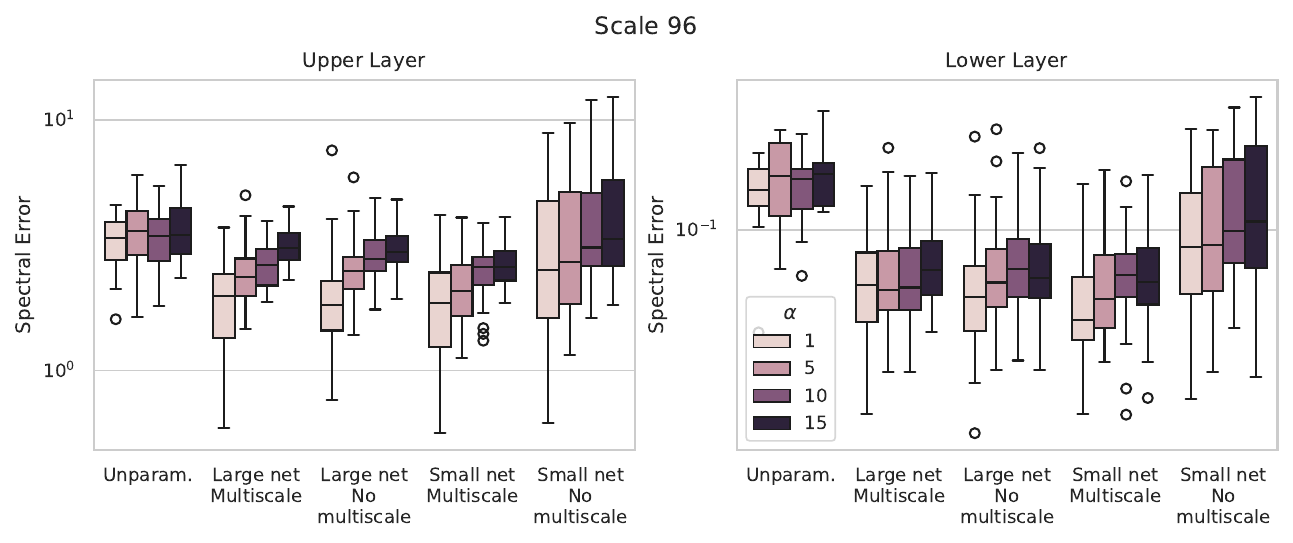}%
    \label{fig:qg-alpha-96}
  \end{subfigure}
  \begin{subfigure}[t]{0.9\linewidth}
    \centering
    \includegraphics[width=\linewidth]{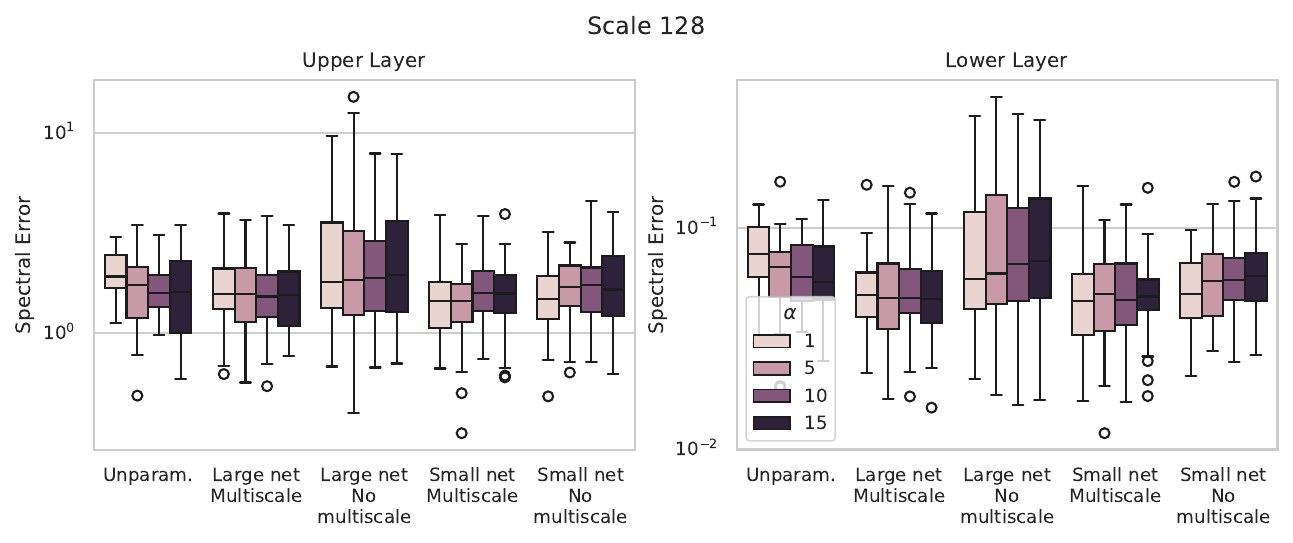}%
    \label{fig:qg-alpha-128}
  \end{subfigure}
  \caption{Spectral errors for trained architectures averaged over
    $5\,400$ steps. Increasing $\alpha$ values reduce the sharpness of
    the QG model's filter, which in turn reduces built-in model
    stability. In each row, the separate panels show values for the
    two layers of the QG system.}%
  \label{fig:qg-alpha}
\end{figure}

To distinguish these cases we also examine the spectral error of these
trajectories, distributions of which are plotted in
Figure~\ref{fig:qg-alpha}. We also run trajectories with varying
values for $\alpha$ which controls the sharpness of the QG model's
internal spectral filter. Higher values of $\alpha$ reduce the
attenuation of higher frequencies, making the QG model less stable and
making errors in the parameterization more evident over time. The
results in this figure suggest that the multiscale training generally
improves model performance. For large architectures on the system at
scale 128, multiscale training reduces instabilities. For the results
on scale 96---where more parameterization is required---adding
multiscale training allows the small architecture to achieve the same
results as the large architecture, allowing for a smaller more
efficient network to yield a more accurate and stable
parameterization.

\subsubsection{Kolmogorov Flow Results}%
\label{sec:comb-nsres}

In addition to our experiments on the QG model we also test our method
on a Kolmogorov Flow (KF) system. Unlike the QG model, the KF system
does not include an internal filter. This results in a simulation
which can be highly unstable, and many of the learned parameterizations
we trained and tested did not successfully complete a stable
trajectory. To reduce this problem we modify our training procedure to
inject Gaussian noise to inputs in order to encourage the learned
parameterizations to be stable to corruptions in the system states.
This noise has mean zero and a configurable scale parameter $\beta$.
Further discussion of our experiments adjusting this parameter are
included later in this section.

\begin{figure}
  \centering
  \begin{subfigure}[t]{0.48\linewidth}
    \centering
    \includegraphics[width=\linewidth]{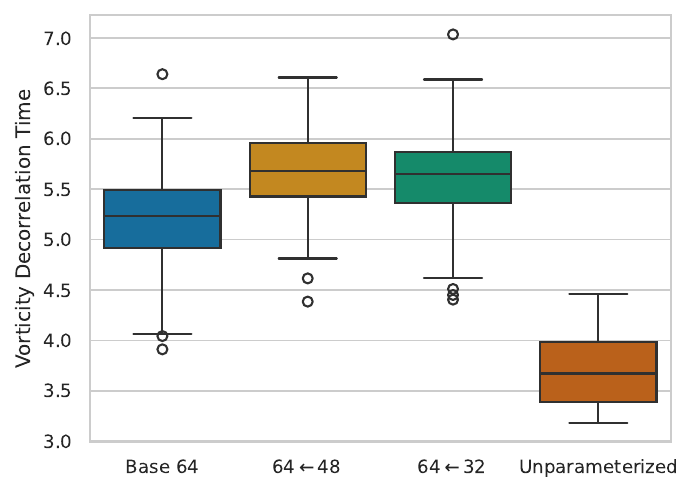}
    \caption{Vorticity decorrelation times from online experiments for the KF system. A longer decorrelation time generally reflects a better parameterization.}%
    \label{fig:ns-decorr-small}
  \end{subfigure}%
  \hspace{0.5em}
  \begin{subfigure}[t]{0.48\linewidth}
    \centering
    \includegraphics[width=\linewidth]{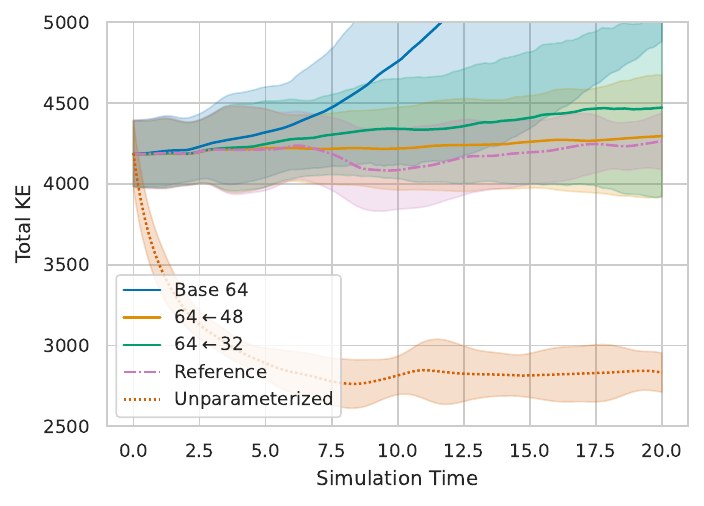}
    \caption{Kinetic energy evolution. Solid lines are means over all
      test trajectories and trained networks and shaded area is a
      $1\sigma$ region.}%
    \label{fig:ns-ke-time-small}
  \end{subfigure}
  \caption{Results of online experiments for Kolmogorov Flow (KF) system. Values are
    computed over a collection of 10 trained networks evaluated on 16
    reference trajectories each.}%
  \label{fig:ns-results}
\end{figure}

Using the best calibrated noise scales, we train a series of small
architecture networks both with multiscale training and with the
single scale baseline architecture. The large architectures displayed
very high instability irrespective of the noise level. As a result
further results on the KF in this section are produced for the small
architecture only. An illustration of this noise calibration issue is
included in Appendix~\ref{sec:extonlres}. For each of these networks
we roll out a series of trajectories online.
Figure~\ref{fig:ns-results} shows the results of this online testing
for the KF experiments. Results are averaged across 10 trained
networks of each type and 16 held out test trajectories for each.

Figure~\ref{fig:ns-decorr-small} reports the observed decorrelation
time in the vorticity channel $\omega$ of the KF system. Results are
in simulation time (out of a full trajectory length of 70.0). Longer
decorrelation times reflect a parameterization which better reproduced
the statistical properties of the reference trajectory even though, due
to the chaos in these systems, the trajectories will eventually
decorrelate. We note that the multiscale runs have a longer
decorrelation time than both the baseline network and the
unparameterized system with the multiscale network using two scales
closer in size having a slightly better performance.

Trends in the evolution of kinetic energy over time are plotted in
Figure~\ref{fig:ns-ke-time-small}. The lines in this figure are the
mean kinetic energy across the test trajectories and networks and the
shaded regions are a $1\sigma$ range. In these results, the multiscale
networks remain more stable over time and show smaller variance at
later time steps. All networks prevent the significant energy loss of
the unparameterized trajectories even though the baseline network
displays significant instability.

The results above were produced using separately chosen values of the
training noise level $\beta$, one noise level for each network
architecture and set of predictino scales. The values of $\beta$ were
set as a fraction of the empirical standard deviation of input values
from the training set. Higher noise levels injected during training
reduce instability but decrease parameterization accuracy, while lower
noise levels fail to correct the problem of instability during online
rollouts.

\begin{figure}
    \centering
    \includegraphics[width=0.9\linewidth]{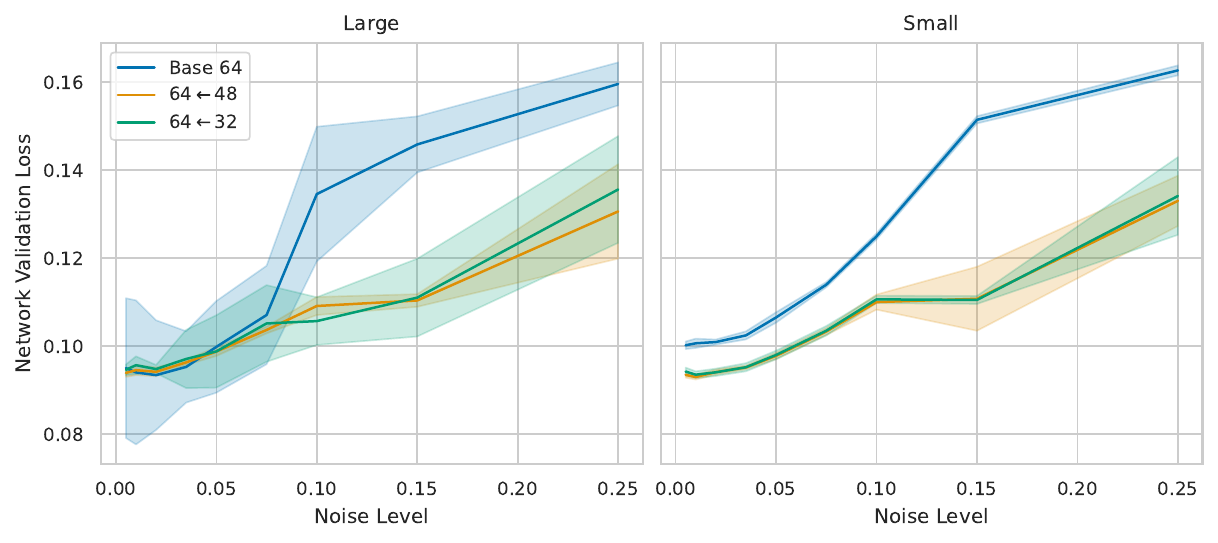}
    \caption{Offline noise calibration for the Kolmogorov Flow (KF)
      system. Solid lines are a mean validation loss observed during
      network training and shaded regions show a $1\sigma$ range. The
      large architecture shows significantly greater variance than the
      small architecture. Increasing noise scale during training
      increases the validation loss but can improve robustness against
      noise during online rollouts.}%
  \label{fig:ns-noise-cal-offline}
\end{figure}

\begin{figure}
    \centering
    \includegraphics[width=0.6\linewidth]{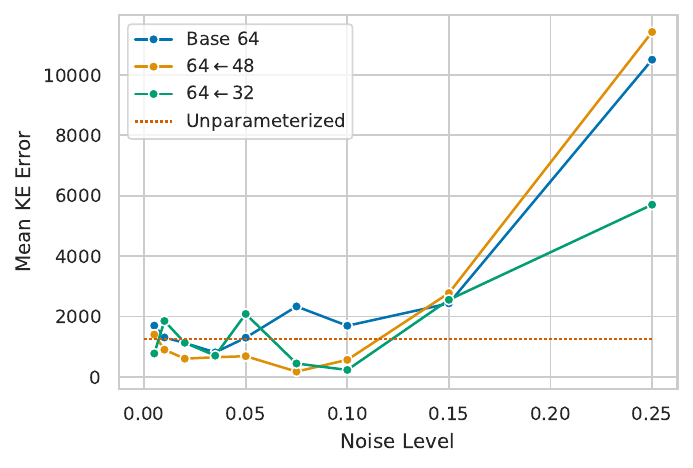}
    \caption{Online noise calibration for the KF system (small
      architecture only). Increasing noise scale eventually reduces
      the networks' ability to correct kinetic energy. We select the
      lowest online kinetic energy error point on each curve for each network as
      the noise scale used during training.}%
  \label{fig:ns-noise-cal-online}
\end{figure}

Figure~\ref{fig:ns-noise-cal-offline} and
Figure~\ref{fig:ns-noise-cal-online} show measurements used in this
calibration. We select our target noise level based in part on the
offline network validation losses and online kinetic energy errors
observed after training. We also show results for a
system scale of size $64$ as larger state sizes did not have
sufficient unresolved dynamics to close on the KF system. Each network
is provided the two velocity components $u$, $v$ as well as a
vorticity $\omega$ computed from these two.

Based on the results in Figure~\ref{fig:ns-noise-cal-offline} we see
that higher noise levels reduce variance in offline validation
quality, but can increase instabilities in online testing as evidenced
by the kinetic energy errors in Figure~\ref{fig:ns-noise-cal-online}.
As a result we selected a noise level of 0.035 for baseline runs,
0.075 for multiscale runs between scales $64\leftarrow48$ and 0.1 for
$64\leftarrow32$ runs (the lowest kinetic energy error point in each curve). These
values were used to train the networks used in the further KF
experiments discussed earlier in this section.

\section{Conclusion}%
\label{sec:conclusion}

Our experiments in this work illustrate the potential advantages
resulting from decomposing the subgrid forcing problem into one across
scales. In particular, our results show that this decomposition
improves stability and accuracy, especially for smaller network
architectures. Such improvements could support the deployment of
machine learning methods to tasks which are constrained by available
computational resources, as is common in climate applications. Our
results show improvements made possible by structuring prediction
tasks to expose important structures of the task. For the fluid
problems considered here, and in other tasks, a multiscale
decomposition is natural and makes use of links between scales in the
model dynamics, and better handles underlying uncertainty in the
parameterization task.

In addition to using a multiscale decomposition in future learned
parameterizations, future work could explore other applications of
this approach. In particular, this decomposition could be advantageous
for stochastic parameterizations; perhaps using a deterministic
downscale prediction as a foundation for later stochastic or
generative outputs or further integrating the multiscale prediction
into the network architecture to realize greater efficiency.

\subsection*{Acknowledgments}%

This research received support through Schmidt Sciences, LLC\@.
This work was also supported by NYU IT High
Performance Computing resources, services, and staff expertise.

\printbibliography{}

@misc{pyqg,
  author       = {Ryan Abernathey and
                  rochanotes and
                  Andrew Ross and
                  Malte Jansen and
                  Ziwei Li and
                  Francis J. Poulin and
                  Navid C. Constantinou and
                  Anirban Sinha and
                  Dhruv Balwada and
                  SalahKouhen and
                  Spencer Jones and
                  Cesar B Rocha and
                  Christopher L. Pitt Wolfe and
                  Chuizheng Meng and
                  Hugo van Kemenade and
                  James Bourbeau and
                  James Penn and
                  Julius Busecke and
                  Mike Bueti and
                  Tobias},
  title        = {pyqg: v0.7.2},
  month        = 5,
  year         = 2022,
  publisher    = {Zenodo},
  version      = {v0.7.2},
  doi          = {10.5281/zenodo.6563667},
  url          = {https://doi.org/10.5281/zenodo.6563667}
}

@misc{jax,
  author = {James Bradbury and Roy Frostig and Peter Hawkins and
                  Matthew James Johnson and Chris Leary and Dougal
                  Maclaurin and George Necula and Adam Paszke and Jake
                  Vander{P}las and Skye Wanderman-{M}ilne and Qiao
                  Zhang},
  title = {{JAX}: composable transformations of {P}ython+{N}um{P}y programs},
  url = {https://github.com/google/jax},
  version = {0.4.26},
  year = 2018,
}

@misc{pyqgjax,
  author       = {Karl Otness},
  title        = {pyqg-jax: v0.8.1},
  month        = 2,
  year         = 2024,
  publisher    = {Zenodo},
  version      = {v0.8.1},
  doi          = {10.5281/zenodo.10719906},
}

@article{equinox,
    author={Patrick Kidger and Cristian Garcia},
    title={{E}quinox: neural networks in {JAX} via callable {P}y{T}rees and filtered transformations},
    year=2021,
    journal={Differentiable Programming workshop at Neural Information Processing Systems 2021},
    doi={10.48550/arXiv.2111.00254},
}

@article{maulik18, title={Subgrid modelling for two-dimensional turbulence using neural networks}, volume={858}, DOI={10.1017/jfm.2018.770}, journal={Journal of Fluid Mechanics}, publisher={Cambridge University Press}, author={Maulik, R. and San, O. and Rasheed, A. and Vedula, P.}, year={2019}, pages={122–144}}

@article{ross23,
author = {Ross, Andrew and Li, Ziwei and Perezhogin, Pavel and Fernandez-Granda, Carlos and Zanna, Laure},
title = {Benchmarking of Machine Learning Ocean Subgrid Parameterizations in an Idealized Model},
journal = {Journal of Advances in Modeling Earth Systems},
volume = {15},
number = {1},
pages = {e2022MS003258},
doi = {10.1029/2022MS003258},
url = {https://agupubs.onlinelibrary.wiley.com/doi/abs/10.1029/2022MS003258},
eprint = {https://agupubs.onlinelibrary.wiley.com/doi/pdf/10.1029/2022MS003258},
note = {e2022MS003258 2022MS003258},
year = {2023}
}

@article{frezat22,
author = {Frezat, Hugo and Le Sommer, Julien and Fablet, Ronan and Balarac, Guillaume and Lguensat, Redouane},
title = {A Posteriori Learning for Quasi-Geostrophic Turbulence Parametrization},
journal = {Journal of Advances in Modeling Earth Systems},
volume = {14},
number = {11},
pages = {e2022MS003124},
doi = {10.1029/2022MS003124},
url = {https://agupubs.onlinelibrary.wiley.com/doi/abs/10.1029/2022MS003124},
eprint = {https://agupubs.onlinelibrary.wiley.com/doi/pdf/10.1029/2022MS003124},
note = {e2022MS003124 2022MS003124},
year = {2022}
}

@article{guillaumin21,
author = {Guillaumin, Arthur P. and Zanna, Laure},
title = {Stochastic-Deep Learning Parameterization of Ocean Momentum Forcing},
journal = {Journal of Advances in Modeling Earth Systems},
volume = {13},
number = {9},
pages = {e2021MS002534},
doi = {10.1029/2021MS002534},
url = {https://agupubs.onlinelibrary.wiley.com/doi/abs/10.1029/2021MS002534},
eprint = {https://agupubs.onlinelibrary.wiley.com/doi/pdf/10.1029/2021MS002534},
note = {e2021MS002534 2021MS002534},
year = {2021}
}

@article{zanna20,
author = {Zanna, Laure and Bolton, Thomas},
title = {Data-Driven Equation Discovery of Ocean Mesoscale Closures},
journal = {Geophysical Research Letters},
volume = 47,
number = 17,
pages = {e2020GL088376},
doi = {10.1029/2020GL088376},
url = {https://agupubs.onlinelibrary.wiley.com/doi/abs/10.1029/2020GL088376},
eprint = {https://agupubs.onlinelibrary.wiley.com/doi/pdf/10.1029/2020GL088376},
note = {e2020GL088376 10.1029/2020GL088376},
year = 2020
}

@article{foxkemper19,
author={Fox-Kemper, Baylor and Adcroft, Alistair and Böning, Claus W. and Chassignet, Eric P. and Curchitser, Enrique and Danabasoglu, Gokhan and Eden, Carsten and England, Matthew H. and Gerdes, Rüdiger and Greatbatch, Richard J. and Griffies, Stephen M. and Hallberg, Robert W. and Hanert, Emmanuel and Heimbach, Patrick and Hewitt, Helene T. and Hill, Christopher N. and Komuro, Yoshiki and Legg, Sonya and Le Sommer, Julien and Masina, Simona and Marsland, Simon J. and Penny, Stephen G. and Qiao, Fangli and Ringler, Todd D. and Treguier, Anne Marie and Tsujino, Hiroyuki and Uotila, Petteri and Yeager, Stephen G.},
title={Challenges and Prospects in Ocean Circulation Models},
journal={Frontiers in Marine Science},
volume=6,
year=2019,
url={https://www.frontiersin.org/articles/10.3389/fmars.2019.00065},
doi={10.3389/fmars.2019.00065},
issn={2296-7745},
}

@article{perezhogin23,
author={Pavel Perezhogin and Laure Zanna and Carlos Fernandez-Granda},
title = {Generative Data-Driven Approaches for Stochastic Subgrid Parameterizations in an Idealized Ocean Model},
journal = {Journal of Advances in Modeling Earth Systems},
volume = 15,
number = 10,
doi = {10.1029/2023MS003681},
year = 2023,
}

@InProceedings{foxkemper14,
  author =       {Baylor Fox-Kemper and Scott Bachman and Brodie Pearson and Scott Reckinger},
  title =        {Principles and advances in subgrid modelling for eddy-rich simulations},
  booktitle = {CLIVAR Exchanges (WGOMD Workshop on High Resolution Ocean Climate Modeling)},
  year =      2014,
  volume =    19,
  pages =     {42--46}}

@article {smagorinsky63,
      author ={J. Smagorinsky},
      title = {General Circulation Experiments with the Primitive Equations: I. The Basic Experiment},
      journal = {Monthly Weather Review},
      year = 1963,
      publisher = {American Meteorological Society},
      volume = 91,
      number = 3,
      doi = "10.1175/1520-0493(1963)091<0099:GCEWTP>2.3.CO;2",
      pages=      {99 - 164},
}

@InProceedings{singer22,
  author =       {Uriel Singer and Adam Polyak and Thomas Hayes and Xi Yin and Jie An and Songyang Zhang and Qiyuan Hu and Harry Yang and Oron Ashual and Oran Gafni and Devi Parikh and Sonal Gupta and Yaniv Taigman},
  title =        {Make-A-Video: Text-to-Video Generation without Text-Video Data},
  booktitle = {ICLR},
  year =      2023,
  doi = {10.48550/arXiv.2209.14792},
}

@Article{ho22,
  author =       {Jonathan Ho and William Chan and Chitwan Saharia and Jay Whang and Ruiqi Gao and Alexey Gritsenko and Diederik P. Kingma and Ben Poole and Mohammad Norouzi and David J. Fleet and Tim Salimans},
  title =        {Imagen Video: High Definition Video Generation with Diffusion Models},
  journal =      {arXiv Preprint},
  year =         2022,
  doi = {10.48550/arXiv.2210.02303},
  }

@misc{dmjax,
  title = {The {D}eep{M}ind {JAX} {E}cosystem},
  author = {Babuschkin, Igor and Baumli, Kate and Bell, Alison and Bhupatiraju, Surya and Bruce, Jake and Buchlovsky, Peter and Budden, David and Cai, Trevor and Clark, Aidan and Danihelka, Ivo and Fantacci, Claudio and Godwin, Jonathan and Jones, Chris and Hemsley, Ross and Hennigan, Tom and Hessel, Matteo and Hou, Shaobo and Kapturowski, Steven and Keck, Thomas and Kemaev, Iurii and King, Michael and Kunesch, Markus and Martens, Lena and Merzic, Hamza and Mikulik, Vladimir and Norman, Tamara and Quan, John and Papamakarios, George and Ring, Roman and Ruiz, Francisco and Sanchez, Alvaro and Schneider, Rosalia and Sezener, Eren and Spencer, Stephen and Srinivasan, Srivatsan and Wang, Luyu and Stokowiec, Wojciech and Viola, Fabio},
  url = {http://github.com/deepmind},
  year = {2020},
}

@inproceedings{kingma14,
  title={Adam: A method for stochastic optimization},
  author={Kingma, Diederik P and Ba, Jimmy},
  year =         2015,
  booktitle = {International Conference on Learning Representations},
  doi={10.48550/arXiv.1412.6980},
}

@article{jansen19,
author = {Jansen, Malte F. and Adcroft, Alistair and Khani, Sina and Kong, Hailu},
title = {Toward an Energetically Consistent, Resolution Aware Parameterization of Ocean Mesoscale Eddies},
journal = {Journal of Advances in Modeling Earth Systems},
volume = {11},
number = {8},
pages = {2844-2860},
doi = {https://doi.org/10.1029/2019MS001750},
url = {https://agupubs.onlinelibrary.wiley.com/doi/abs/10.1029/2019MS001750},
year = {2019}
}

@article{jaxcfd,
  author = {Kochkov, Dmitrii and Smith, Jamie A. and Alieva, Ayya and Wang, Qing and Brenner, Michael P. and Hoyer, Stephan},
  title = {Machine learning{\textendash}accelerated computational fluid dynamics},
  volume = 118,
  number = 21,
  elocation-id = {e2101784118},
  year = 2021,
  doi = {10.1073/pnas.2101784118},
  publisher = {National Academy of Sciences},
  issn = {0027-8424},
  URL = {https://www.pnas.org/content/118/21/e2101784118},
  eprint = {https://www.pnas.org/content/118/21/e2101784118.full.pdf},
  journal = {Proceedings of the National Academy of Sciences}
}

@InProceedings{song21,
  title={Score-Based Generative Modeling through Stochastic Differential Equations},
  author={Yang Song and Jascha Sohl-Dickstein and Diederik P. Kingma and Abhishek Kumar and Stefano Ermon and Ben Poole},
  booktitle = {ICLR},
  year =      2021,
  doi = {10.48550/arXiv.2011.13456},
}

@article{yang22,
  title={Diffusion Models: A Comprehensive Survey of Methods and Applications},
  author={Ling Yang and Zhilong Zhang and Yang Song and Shenda Hong and Runsheng Xu and Yue Zhao and Wentao Zhang and Bin Cui and Ming-Hsuan Yang},
  year = {2023},
  publisher = {Association for Computing Machinery},
  volume = {56},
  number = {4},
  doi = {10.1145/3626235},
  journal = {ACM Computing Surveys},
  month = {11},
}

\appendix
\section{Models}%
\label{sec:modelextdef}

In this work we consider two models: a quasi-geostrophic model (QG)
and a Kolmogorov Flow model (KF). Further information on these systems
is provided below.

\subsection{Quasi-Geostrophic Model}%
\label{sec:qgdef}

For our experiments we target the two-layer quasi-geostrophic model
implemented in PyQG which is a simplified approximation of
fluid dynamics~\cite{pyqg}. This model follows the evolution of a
potential vorticity $q$, divided into two layers $q = [q_1, q_2]$.
This system is pseudo-spectral and has periodic boundaries along the
edges of each layer. The evolution of the quantities in
Fourier space (indicated by a hat) is:
\begin{align}
  \label{eq:qgevol}
  \frac{\partial{}\hat{q}_1}{\partial{}t} &= -\widehat{J(\psi_1, q_1)} - i k \beta_1\hat{\psi}_1 + \widehat{\text{ssd}}\\
  \frac{\partial{}\hat{q}_2}{\partial{}t} &= -\widehat{J(\psi_2, q_2)} - i k \beta_2 \hat{\psi}_2 + r_{\text{ek}}\kappa^2\hat{\psi}_2 + \widehat{\text{ssd}}
\end{align}
where $J(A, B) \defeq A_{x}B_{y} - A_{y}B_{x}$, ``$\text{ssd}$'' is a small scale
dissipation, and the quantity $\psi$ is related to $q$ by:
\begin{equation}
  \label{eq:2}
  \begin{bmatrix}
    -(\kappa^2 + F_1) & F_1\\
    F_2 & -(\kappa^2 + F_2)
  \end{bmatrix}
  \begin{bmatrix}
    \hat{\psi}_1 \\ \hat{\psi}_2
  \end{bmatrix}
  =
  \begin{bmatrix}
    \hat{q}_1 \\ \hat{q}_2
  \end{bmatrix}.
\end{equation}
The values $\kappa$ are the radial wavenumbers $\sqrt{k^2 + l^2}$
while $k$ and $l$ are wavenumbers in the zonal and meridional
directions (the axis-aligned directions in our grid), respectively~\cite{ross23}.

We use the ``eddy'' configuration from~\cite{ross23} which sets the
following values for model constants:
\begin{equation*}
  \begin{aligned}[t]
    r_{\text{ek}} &= 5.787 \times 10^{-7}\\
    \delta &= \frac{H_1}{H_2} = 0.25\\
    \beta &= 1.5\times 10^{-11}\\
    r_d &= 15\,000
  \end{aligned}
  \hspace{1in}
  \begin{aligned}[t]
    F_1 &= \frac{1}{r_d^{2}(1 + \delta)}\\
    F_2 &= \delta F_1\\
    W &= 10^{6}\\
    L &= 10^{6}
  \end{aligned}
\end{equation*}
where $H_1, H_2$ are the heights of each of the two layers of $q$ and
$r_d$ is a deformation radius. For more information on the model
configuration, consult~\cite{ross23} and the documentation for the
PyQG package.

We generate our data at a ``true'' resolution on a grid of dimension
$256 \times 256$ using the PyQG default third order
Adams-Bashforth method for time stepping. We use a time step of
$\Delta t = 3\,600$ generating $86\,400$ steps from which we keep every eighth
leaving $10\,800$ per trajectory. Our training set consists of 100 such
trajectories, and our evaluation set contains 10.

Each step produces a ground truth potential vorticity
$q_{\text{true}}$ along with a spectral time derivative
$\partial{}\hat{q}_{\text{true}}/\partial{}t$. From these we apply our family of
coarsening operators $C$ (described in Appendix~\ref{sec:coarsedef}) to
produce filtered and coarsened values $q_{\text{lr}} \defeq C_{\text{lr}}(q_{\text{true}})$ at resolutions of
$128\times 128$, $96 \times 96$, and $64 \times 64$.

For each of these, we recompute spectral time derivatives in a
coarsened PyQG model $\partial{}\hat{q}_{\text{lr}}/\partial{}t$, and we pass each
time derivative to spatial variables and compute the target forcing
for this scale:
\begin{equation*}
  S_{\text{lr}} = C_{\text{lr}}\dparen[\Big]{\frac{\partial{}q_{\text{true}}}{\partial{}t}} - \frac{\partial{}q_{\text{lr}}}{\partial{}t}.
\end{equation*}
These forcings---at each of the three scales---along with the high
resolution variables are stored in the training and evaluation sets for
each step.

\subsection{Kolmogorov Flow Model}%
\label{sec:nsdef}

We also consider a Kolmogorov Flow system as implemented by the
JAX-CFD package~\cite{jaxcfd}. This is a configuration of a
Navier-Stokes system with a forcing of $\sin(4y)\hat{x} - 0.1u$ where
the subtracted term is a small velocity-dependent drag.

We generate our ground truth data at a true resolution of
$2048\times2048$ on a domain of size $4\pi$ on each side (double the
default size of $2\pi$). We configure the system to have a viscosity
of $1/3500$ which, with the domain, produces a Reynolds number of
$7\,000$. States are ported to lower resolutions using the
\texttt{downsample\_staggered\_velocity} function in the JAX-CFD
package.

\section{Network Architecture and Training}%
\label{sec:nndef}

The network architectures used in this paper are all feedforward
convolutional neural networks. The structure and training parameters
of each network vary with the target system and were adjusted for the
online combined experiments using the results of the offline separated
experiments.

\subsection{Separated Experiment Architectures}%
\label{sec:nndef-sep}

For our \emph{separated} experiments on the QG system we use two
different feedforward CNN architectures from previous work without
batch norm~\cite{guillaumin21}. We take the architectural parameters
from this work as our default ``small'' architecture, while the
``large'' architecture for these experiments roughly doubles the size
of each convolution kernel. This produces the architectures listed in
Table~\ref{tab:cnndef}. We use $\mathrm{ReLU}$ activations between
each convolution. Each convolution is performed with periodic padding,
matching the boundary conditions of the system. All convolutions are
with bias. The input and output channel counts are determined by the
inputs of the network. For the QG system each input has two layers,
each of which is handled as a separate channel. Quantities for the KF
system have only a single layer each. These parameters are adjusted
for each task to accommodate the inputs and make the required
predictions. We implement our networks with Equinox~\cite{equinox}.

\begin{table}[ht]
  \centering
  \begin{tabular}{lrrr}
    \toprule
    Conv. Layer & Chans. Out & Small Kernel Size & Large Kernel Size\\
    \midrule
    1 & 128 & $(5, 5)$ & $(9, 9)$\\
    2 & 64 & $(5, 5)$ & $(9, 9)$\\
    3 & 32 & $(3, 3)$ & $(5, 5)$\\
    4 & 32 & $(3, 3)$ & $(5, 5)$\\
    5 & 32 & $(3, 3)$ & $(5, 5)$\\
    6 & 32 & $(3, 3)$ & $(5, 5)$\\
    7 & 32 & $(3, 3)$ & $(5, 5)$\\
    8 & out layers & $(3, 3)$ & $(5, 5)$\\
    \bottomrule
  \end{tabular}
  \caption{Architecture specifications for each neural network used in
    the separated experiments. Convolution kernel sizes vary between
    the architecture sizes. The channel counts are adjusted to
    accommodate the inputs and outputs of each task.}%
  \label{tab:cnndef}%
\end{table}

We train each network with the Adam optimizer~\cite{kingma14} as
implemented in Optax~\cite{dmjax}. The learning rate is set to a
constant depending on architecture size: the small networks use
$5\times10^{-4}$, while the large networks use $2\times10^{-4}$. The
networks are trained to minimize MSE loss. Large chunks of 10\,850
steps are sampled with replacement from the dataset which is
pre-shuffled uniformly. Then each of these chunks is shuffled again
and divided into batches of size 256 without replacement. One epoch
consists of 333 such batches. We train the small networks for 132
epochs, and the large networks for 96 epochs. We store the network
weights which produced the lowest training set loss and use these for
evaluation.

For all input and target data, we compute empirical means and standard
deviations and standardize the overall distributions by these values
before passing them to the network. The means and standard deviations
from the training set are used in evaluation as well.

\subsection{Combined Experiment Architectures}%
\label{sec:nndef-comb}

For our combined experiments, which carry out online tests on both the
QG and KF systems, we made some modifications to the tested
architectures in order to explore the efficiency improvements which
could be realized by our multiscale approach. We kept the same
``large'' architecture as was used in the separated experiments, but
carried out an architecture search to select a ``small'' architecture.

Because each overall parameterization combines two sub-networks (see
Figure~\ref{fig:combillust}) we describe each network by giving a
description of the architectures of each component network. These are
described by the sizes of the convolution kernels (``pure small
(psm),'' ``small (sm),'' ``pure medium (pmd),'' and ``medium (md),''
respectively) and the number of convolution layers, either 4 or 8. The
``large'' architecture described in Table~\ref{tab:cnndef} would be
described as ``md8'' and the combination of two of these is ``md8md8.''

\begin{figure}
  \centering
  \includegraphics[width=0.95\linewidth]{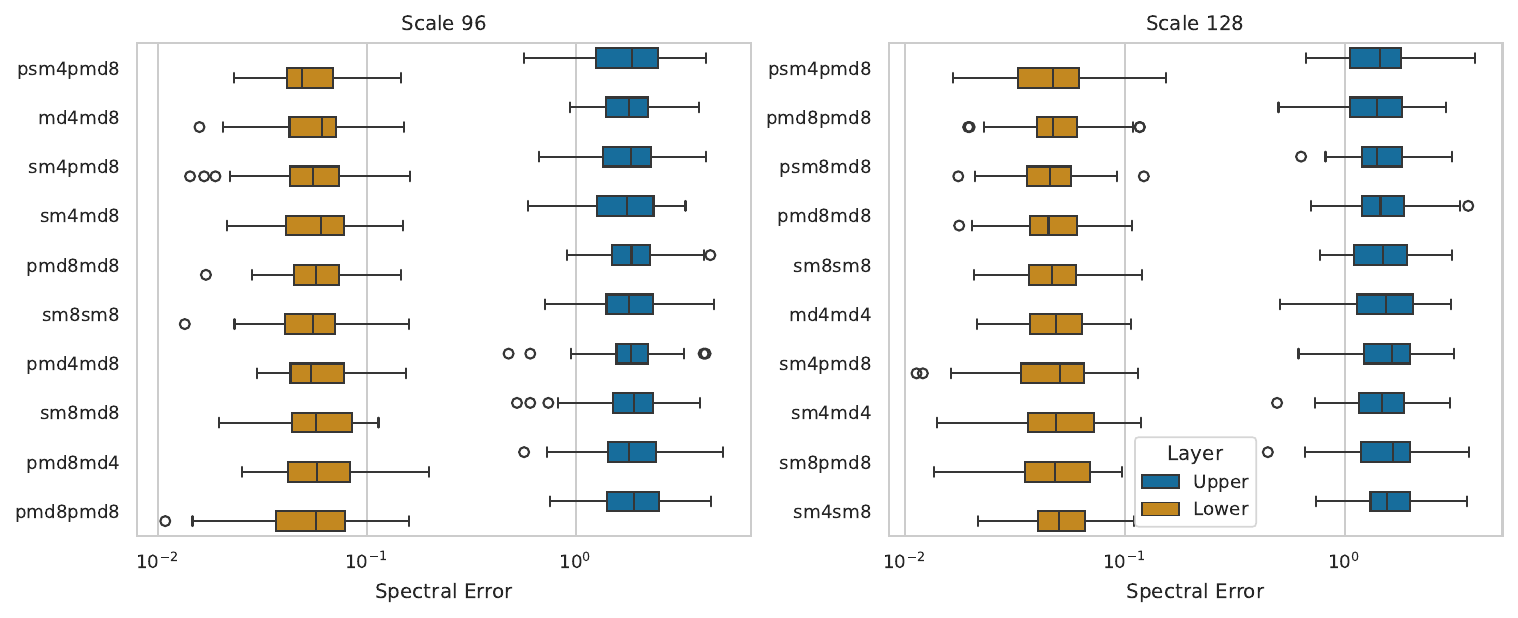}
  \caption{Architecture selection by scale size on QG system for the
    combined experiments.}%
  \label{fig:arch-select}
\end{figure}

Using these options, four networks of each architecture were trained
on the QG system and tested online for the resulting spectral errors.
The results of each of the two network scales are reported in
Figure~\ref{fig:arch-select}. These were ranked by increased spectral
errors separately for each layer, then each architecture choice was
ranked by the worst of these two layer ranks.

\begin{table}[ht]
  \centering
  \begin{tabular}{lrrrrrr}
    \toprule
    \multirow{2}{*}{Conv. Layer} & \multicolumn{2}{c}{md8} & \multicolumn{2}{c}{psm4} & \multicolumn{2}{c}{pmd8} \\
    \cmidrule(r){2-3}     \cmidrule(lr){4-5} \cmidrule(l){6-7}
                                 &Chans.\ out&Kernel&Chans.\ out&Kernel&Chans.\ out&Kernel\\
    \midrule
    1 & 128 & $(9,9)$ & 128 & $(3,3)$ & 128 & $(5,5)$\\
    2 & 64 & $(9,9)$ & 64 & $(3,3)$ & 64 & $(5,5)$\\
    3 & 32 & $(5,5)$ & 32 & $(3,3)$ & 32 & $(5,5)$\\
    4 & 32 & $(5,5)$ & out layers & $(3,3)$ & 32 & $(5,5)$\\
    5 & 32 & $(5,5)$ &  &  & 32 & $(5,5)$\\
    6 & 32 & $(5,5)$ &  &  & 32 & $(5,5)$\\
    7 & 32 & $(5,5)$ &  &  & 32 & $(5,5)$\\
    8 & out layers & $(5,5)$ &  &  & out layers & $(5,5)$\\
    \bottomrule
  \end{tabular}
  \caption{Architecture specifications for each neural network used in
    the combined experiments including those selected through the
    architecture search.}%
  \label{tab:cnndef-archsearch}
\end{table}

The winning ``small'' architecture for both QG scales was
``psm4pmd8.'' Details of these parameters are provided in
Table~\ref{tab:cnndef-archsearch}.

The training parameters also varied from the separated experiments due
to the new end-to-end training configuration. For the QG system
training was conducted using the Adam optimizer following a cosine
annealing schedule with one epoch of linear learning rate warmup. The
``md'' and ``pmd'' networks were trained for 50 epochs at a learning
rate of 0.0004 while the ``psm'' network was trained for 100 epochs
(with 374 batches per epoch) with a learning rate of 0.001. All
batches had size 64.

For the KF system, training was carried out using the Adam optimizer
with $\epsilon=0.001$ (modified from the default). These networks
followed the same cosine annealing with warmup schedule as the QG
systems but had an ending learning rate of 0.0001 and a peak learning
rate of $7.5\times10^{-4}$ for 150 epochs with batches of size 32.
Each epoch consisted of 374 batches. Each training run selected the
network with the best validation loss.

\section{Coarsening Operators}%
\label{sec:coarsedef}

In this work we make use of two families of coarsening operators to
transform system states across scales. The first, denoted $C$, is used
when generating our data. This operator is applied to the ``true''
resolution system outputs $q_{\text{true}}$ and $\partial{}q_{\text{true}}/\partial{}t$
to produce training and evaluation set samples as well as target
forcings $S$. The second operator $D$ (with associated upscaling
$D^{+}$) is applied as a part of each prediction task to adjust scales
around the neural networks as needed. These are the operators
referenced in Figure~\ref{fig:daillust} and Figure~\ref{fig:bdillust}.

Each of these operators is built around a core spectral truncation
operation, $\mathcal{D}$. For an input resolution $\text{hr}$ and an
output resolution $\text{lr}$, this operator truncates the 2D-Fourier
spectrum to the wavenumbers which are resolved at the output
resolution, then spatially resamples the resulting signal for the
target size $\text{lr}$. These operators also apply a scalar
multiplication to adjust the range of the coarsened values. We define
a ratio $\rho \defeq \text{hr}/\text{lr}$.

\subsection{Data Filtering}

The data filtering operator $C$ for the QG system is ``Operator~1'' as
described in~\cite{ross23}. It is a combination of the truncation
operator $\mathcal{D}$ with a spectral filter $\mathcal{F}$
\begin{equation*}
  C \defeq \rho^{-2} \cdot \mathcal{F} \circ \mathcal{D}
\end{equation*}
where the filter $\mathcal{F}$ acts on the 2D-Fourier spectrum of the
truncated value. $\mathcal{F}$ is defined in terms of the radial
wavenumber $\kappa = \sqrt{k^2 + l^2}$ where $k$ and $l$ are the
wavenumbers in each of the two dimensions of the input. For an input
$\hat{v}_{\kappa}$ at radial wavenumber $\kappa$ we define:
\begin{equation*}
  \mathcal{F}(\hat{v}_{\kappa}) = \begin{cases}
    \hat{v}_{\kappa} & \text{if $\kappa \leq \kappa^c$}\\
    \hat{v}_{\kappa} \cdot e^{-23.6\alpha{(\kappa-\kappa^{c})}^{4}\Delta x^4_{\text{lr}}} & \text{if $\kappa > \kappa^c$}
  \end{cases}
\end{equation*}
where $\Delta x_{\text{lr}} \defeq L/\text{lr}$ ($L$ is a system
parameter; see Appendix~\ref{sec:qgdef} for details), and
$\kappa^{c} \defeq (0.65 \pi)/\Delta x_{\text{lr}}$ is a cutoff
wavenumber where decay begins.

For the KF system, we used the routine\\
\begin{minipage}{1.0\linewidth}
  \centering
  \texttt{jax\_cfd.base.resize.downsample\_staggered\_velocity}
\end{minipage}\\
from the JAX-CFD package which computes means of the velocity values
along a specific face of control volumes composed of groups of grid
squares.

\subsection{Rescaling Operator}

For scale manipulations as part of our learned model we make use of a
scaled spectral truncation operator. We define a downscaling operator
$D$ as well as an upscaling operator $D^{+}$:
\begin{equation}
  \label{eq:truncdef}
  D \defeq \rho^{-2}\mathcal{D} \qquad\text{and}\qquad D^+ \defeq \rho^{2}\mathcal{D}^T.
\end{equation}
Note that $D^{+}$ is a right side inverse $DD^{+} = I$, and that
$D^{+}$ is the pseudoinverse $D^{+} = D{(DD^{T})}^{-1}$ because
$\mathcal{D}\mathcal{D}^T=I$. This operator omits the filtering
$\mathcal{F}$ performed as part of coarsening operator $C$ to avoid
numerical issues when inverting the additional spectral filtering.
This operator was used with both QG and KF systems.

\section{Extended Online Results}%
\label{sec:extonlres}

\begin{figure}
  \centering
  \includegraphics[width=0.9\linewidth]{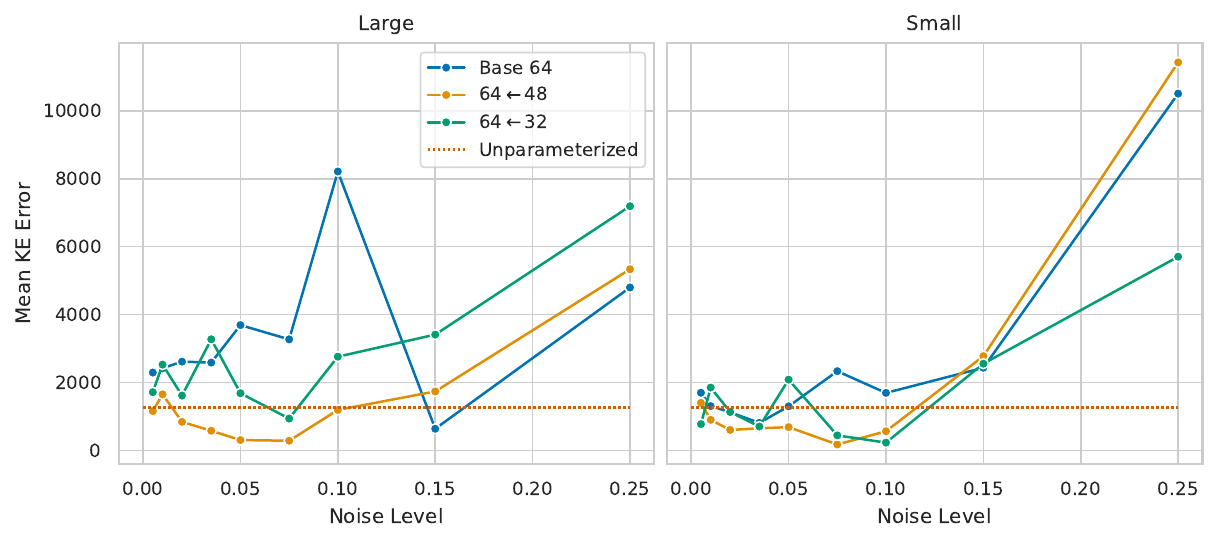}
  \caption{An extended version of Figure~\ref{fig:ns-noise-cal-online}
    including results for the large network architecture. These
    architectures show significant instability, particularly for the
    baselines, even with significant added noise.}%
  \label{fig:ns-noise-cal-online-ext}
\end{figure}

\begin{figure}
  \centering
  \includegraphics[width=0.9\linewidth]{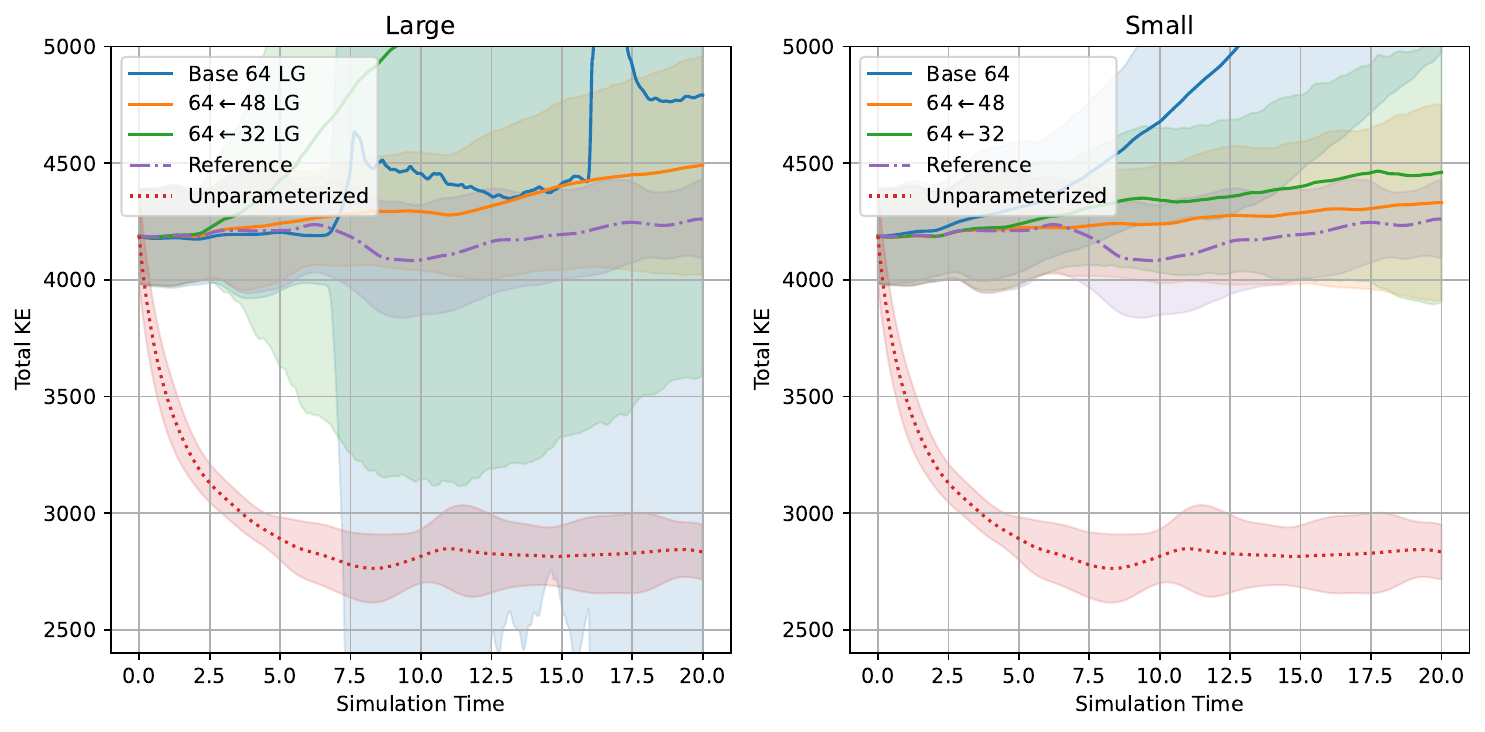}
  \caption{An extended version of Figure~\ref{fig:ns-ke-time-small}
    including results for the large network architecture which
    produced generally unstable results even with attempts to improve
    stability. However even here it appears that the added information
    from a scale of size 48 may help improve network stability.}%
  \label{fig:ns-ke-time-ext}
\end{figure}

Extended KF calibration results for large architectures are presented
in Figure~\ref{fig:ns-noise-cal-online-ext} and
Figure~\ref{fig:ns-ke-time-ext}. The large architectures had
persistent stability problems on the KF system that could not be
resolved with added noise during training.

\end{document}